\documentclass[twocolumn]{aastex631}
\usepackage{xspace}
\newcommand{\mum}{$\mu$m\xspace}
\newcommand{\cmc}{cm$^{-3}$\xspace}

\newcommand{\sii}{[\ion{S}{2}]\xspace}
\newcommand{\Spit}{\emph{Spitzer}\xspace}

\newcommand{\oiii}{[\ion{O}{3}]\xspace}
\newcommand{\ha}{H$\alpha$\xspace}
\newcommand{\hb}{H$\beta$\xspace}

\newcommand{\siidoublet}{\sii:$\lambda$6717/6731\xspace}
\newcommand{\siiha}{[\ion{S}{2}]/H$\alpha$\xspace}

\newcommand{\Msun}{M$_{\odot}$\xspace}
\newcommand{\Msunpc}{M$_{\odot}$ pc$^{-2}$\xspace}

\usepackage{pifont}
\let\tablenum\relax
\usepackage{siunitx}
\usepackage{xcolor}

\shorttitle{JWST observations of M33 SNRs}
\shortauthors{Sarbadhicary et al.}

\begin{document}

\title{A First-look at Spatially-resolved Infrared Supernova remnants in M33 with JWST}

\author[0000-0002-4781-7291]{Sumit K. Sarbadhicary}
\affiliation{Department of Physics, The Ohio State University, Columbus, Ohio 43210, USA}
\affiliation{Center for Cosmology \& Astro-Particle Physics, The Ohio State University, Columbus, Ohio 43210, USA}
\affiliation{Department of Astronomy, The Ohio State University, Columbus, Ohio 43210, USA}

\author[0000-0002-5204-2259]{Erik~Rosolowsky}
\affiliation{Department of Physics, University of Alberta, Edmonton, AB T6G 2E1, Canada}

\author[0000-0002-2545-1700]{Adam K. Leroy}
\affiliation{Center for Cosmology \& Astro-Particle Physics, The Ohio State University, Columbus, Ohio 43210, USA}
\affiliation{Department of Astronomy, The Ohio State University, Columbus, Ohio 43210, USA}

\author[0000-0002-0012-2142]{Thomas G. Williams}
\affiliation{Sub-department of Astrophysics, Department of Physics, University of Oxford, Keble Road, Oxford OX1 3RH, UK}

\author[0000-0001-9605-780X]{Eric W. Koch}
\affiliation{Center for Astrophysics $\mid$ Harvard \& Smithsonian, 60 Garden St., 02138 Cambridge, MA, USA}

\author[0000-0002-5937-9778]{Joshua Peltonen}
\affiliation{Department of Physics, University of Alberta, Edmonton, AB T6G 2E1, Canada}

\author[0000-0003-2599-7524]{Adam Smercina}\thanks{Hubble Fellow}
\affiliation{Space Telescope Science Institute, 3700 San Martin Dr., Baltimore, MD 21218, USA}
\affiliation{Department of Astronomy, Box 351580, University of Washington, Seattle, WA 98195, USA}

\author[0000-0002-1264-2006]{Julianne J.\ Dalcanton}
\affiliation{Center for Computational Astrophysics, Flatiron Institute, 162 Fifth Avenue, New York, NY 10010, USA}
\affiliation{Department of Astronomy, Box 351580, University of Washington, Seattle, WA 98195, USA}

\author[0000-0001-6708-1317]{Simon C. O. Glover}
\affiliation{Universit\"at Heidelberg, Zentrum fur Astronomie, Institut fur Theoretische Astrophysik, Albert-Ueberle-Str. 2, 69120 Heidelberg, Germany}

\author[0000-0003-3252-352X]{Margaret Lazzarini}
\affiliation{Department of Physics \& Astronomy, California State University Los Angeles, 5151 State University Drive, Los Angeles, CA 90032, USA}

\author[0000-0001-8241-7704]{Ryan~Chown}
\affiliation{Department of Astronomy, The Ohio State University, Columbus, Ohio 43210, USA}

\author[0000-0002-3106-7676]{Jennifer Donovan Meyer}
\affiliation{National Radio Astronomy Observatory, 520 Edgemont Rd., Charlottesville, VA 22903, USA}

\author[0000-0002-4378-8534]{Karin Sandstrom}
\affil{Department of Astronomy \& Astrophysics, University of California, San Diego, 9500 Gilman Drive, La Jolla, CA 92093, USA}

\author[0000-0002-7502-0597]{Benjamin F. Williams}
\affiliation{Department of Astronomy, Box 351580, University of Washington, Seattle, WA 98195, USA}

\author[0000-0003-1356-1096]{Elizabeth Tarantino}
\affil{Space Telescope Science Institute, 3700 San Martin Drive, Baltimore, MD 21218, USA}


\correspondingauthor{Sumit K. Sarbadhicary}
\email{sarbadhicary.1@osu.edu}

\begin{abstract}
We present the first spatially-resolved infrared images of supernova remnants (SNRs) in M33 with the unprecedented sensitivity and resolution of JWST. We analyze 40 SNRs in four JWST fields: two covering central and southern M33 with separate NIRCam (F335M, F444W) and MIRI (F560W, F2100W) observations, one $\sim$5 kpc-long radial strip observed with MIRI F770W, and one covering the giant HII region NGC 604 with multiple NIRCam and MIRI broad/narrowband filters. Of the 21 SNRs in the MIRI (F560W+F2100W) field, we found three clear detections (i.e., identical infrared and \ha morphologies), and six partial-detections, implying a detection fraction of 43\% in these bands. One of the SNRs in this field, L10-080, is a potential candidate for having freshly-formed ejecta dust, based on its size and centrally-concentrated 21 \mum emission. In contrast, only one SNR (out of 16) is detectable in the NIRCam F335M+F444W field. Two SNRs near NGC 604 have strong evidence of molecular (H$_2$) emission at 4.7 \mum, making them the farthest known SNRs with visible molecular shocks. Five SNRs have F770W observations, with the smaller younger objects showing tentative signs of emission, while the older, larger ones have voids. Multi-wavelength data indicate that the clearly-detected SNRs are also among the smallest, brightest at other wavelengths (\ha, radio and X-ray), have the broadest line widths (H$\alpha$ FWHM$\sim$250-350 km/s), and the densest environments. No correlation between the JWST-detectability and local star-formation history of the SNRs is apparent. 

\end{abstract}

\keywords{Supernova remnants (1667) --- Interstellar medium (847) --- Shocks (2086) --- James Webb Space Telescope (2291)}


\section{Introduction}
Infrared (IR) studies of supernova remnants (SNRs) provide a unique window into the physics of SN explosions and the interaction of their blastwaves with the ambient medium. Mid-to-far-IR ($>$20 \mum) emission of SNR is generally dominated by thermal emission from collisionally-heated dust grains with a range of sizes and temperatures. Near-to-mid-IR (1-20 \mum) emission in SNRs features a rich tapestry of line emission from shock-excited atomic and molecular species, prominent broad emission features from  polycyclic aromatic hydrocarbons (PAHs), and in rare instances, synchrotron emission \citep[see][for reviews]{Dwek1992, Reach2006, Tielens2008,Williams2017, Vink2020}. 

Both of these observational regimes are now accessible with JWST, and offer valuable insight into SNRs and ISM physics. The mid-to-far-IR observations have been crucial for estimates of the dust mass, composition, formation and destruction rates in the ejecta of younger SNRs \citep[e.g.,][]{Stani2005, Borkowski2006, Williams2006, Sand2009, Rho2009, Temim2013} and in the ISM around older SNRs \citep[e.g.,][]{Temim2015, Laki2015, Seok2015, Chawner2019, Chawner2020, Matsuura2022}. These estimates have been critical to our understanding of how dust formed in the universe \citep[see][and references therein]{Sarangi2018, Micelotta2018}, and in many cases have also helped constrain the unknown explosion mechanisms and progenitors of SNRs \citep[see][and references therein]{Williams2017}. Near-IR spectra have provided rich insights into how SN shocks impact and alter the conditions of ambient molecular clouds \citep[e.g.,][]{Oliva1989, Reach2005, Reach2006, Koo2023}. The presence of these shocked near-IR lines offers some of the most direct evidence that a non-negligible fraction of SNe interact with dense molecular clouds, which is relevant to the efficacy of stellar feedback models \citep{Iffrig2015, Walch2015, Mayker2023,Sarbadhicary2023} and cosmic ray acceleration \citep[e.g.,][]{Slane2015, Sano2021}.
\begin{figure}
    \centering
    \includegraphics[width=\columnwidth]{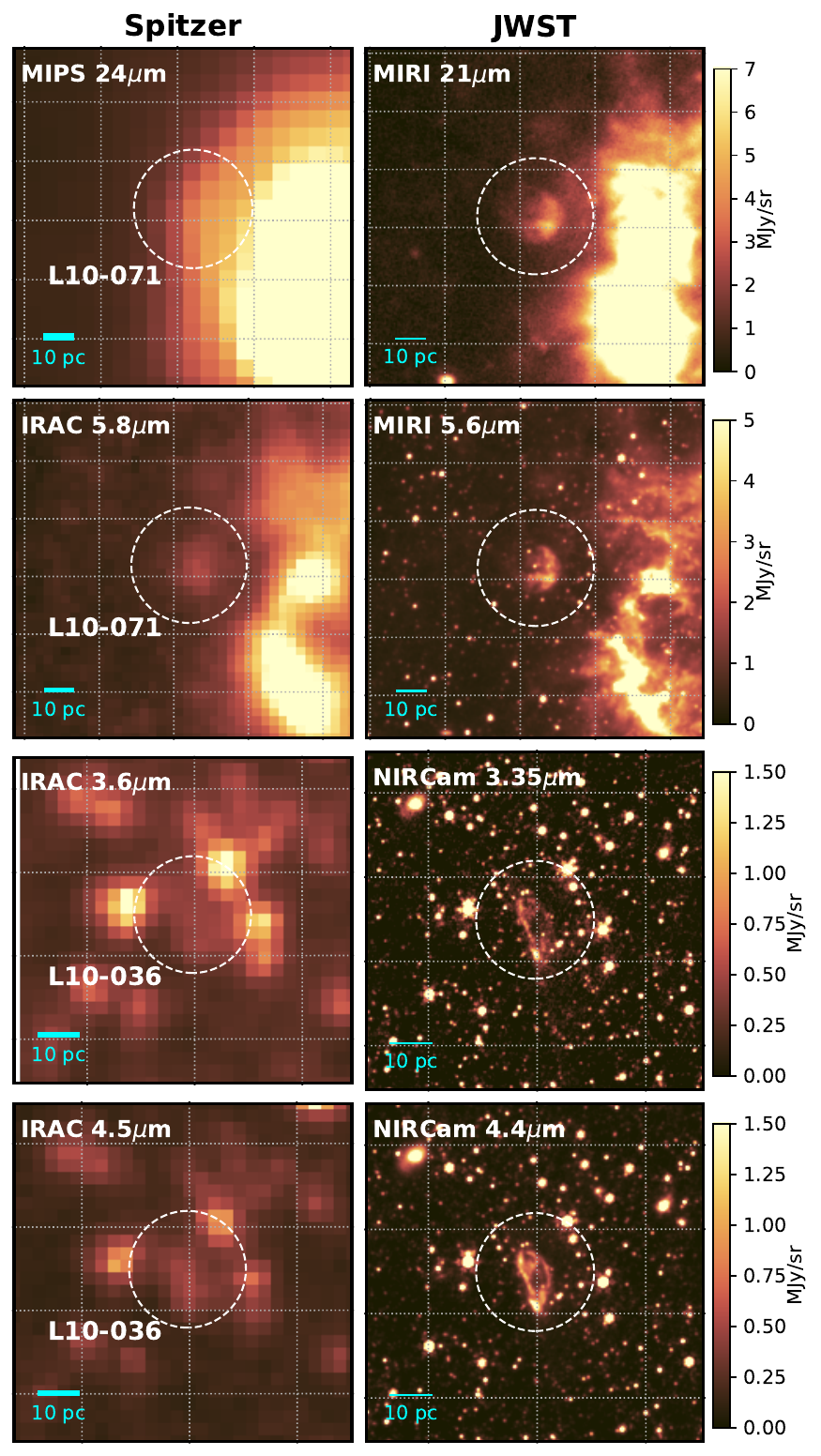}
    \caption{JWST images of SNRs (right column) reveal unprecedented detail compared to the previous generation \Spit (left column). The top two rows show a comparison between \Spit IRAC/MIPS bands and equivalent MIRI bands for SNR L10-071, one of the bright SNRs in the MIRI field. The bottom two rows compare the \Spit IRAC bands with equivalent NIRCam bands for the bright SNR L10-036 from our Center field. Each row shares the same colorbar.}
    \label{fig:jwst-spitzer-comparison}
\end{figure}

Unfortunately, these valuable IR studies of SNRs have thus far been restricted to the Milky Way and the Magellanic Clouds due to the limits of previous-generation IR observatories. The first generation IR studies of SNRs were done with the Infrared Astronomical Satellite (IRAS) and Infrared Space Observatory (ISO) which, in spite of the coarse ($\sim$arcminute) spatial resolution and severe confusion in the Galactic plane, found emission associated with about 30\% of the Galactic SNRs between 10-100 \mum \citep[e.g.][]{Arendt1989, Saken1992, Douvion2001}. 

The observational landscape improved significantly with the launch of \Spit and \emph{Herschel}, whose higher angular resolution across a much wider coverage of IR wavelengths (2\arcsec$-$40\arcsec\  between 3$-$160 \mum for \Spit and 6\arcsec$-$36\arcsec\ between 70$-$500 \mum for Herschel) enabled better detection and cleaner separation of SNR-related emission from confusing sources. These capabilities led to more accurate constraints on dust masses, along with composition, densities, temperatures, and kinematics of shocked clouds in the ejecta and ISM. These are exemplified in many subsequent \Spit and \emph{Herschel} studies of Galactic SNRs \citep[e.g.][]{Reach2006, Blair2007, Rho2008, Barlow2010, Pinheiro2011, Gomez2012, Temim2017, Chawner2019, Chawner2020}, especially leveraging wide-field Galactic plane surveys such as the Galactic Legacy Infrared Midplane Survey Extraordinaire survey \citep[GLIMPSE,][]{GLIMPSE},  MIPS Galactic Plane
Survey \citep[MIPSGAL,][]{MIPSGAL} and Herschel infrared Galactic Plane Survey \citep[Hi-GAL,][]{HIGAL}. 

The increased angular resolution of these observatories also led to the first high-quality extragalactic IR SNR surveys, which can circumvent the confusion and distance uncertainties that are present in Galactic SNR studies. These spatially resolved extragalactic surveys were confined to SNRs in the Magellanic Clouds \citep{Williams2006b, Seok2013, Laki2015, Temim2015, Matsuura2022}. While attempts have been made to extend such studies to SNRs in even more distant galaxies \citep[e.g. M31,][]{Wang2022}, the resolution and sensitivity of \Spit and \emph{Herschel} significantly limited such work.
\begin{figure}
    \centering
    \includegraphics[width=\columnwidth]{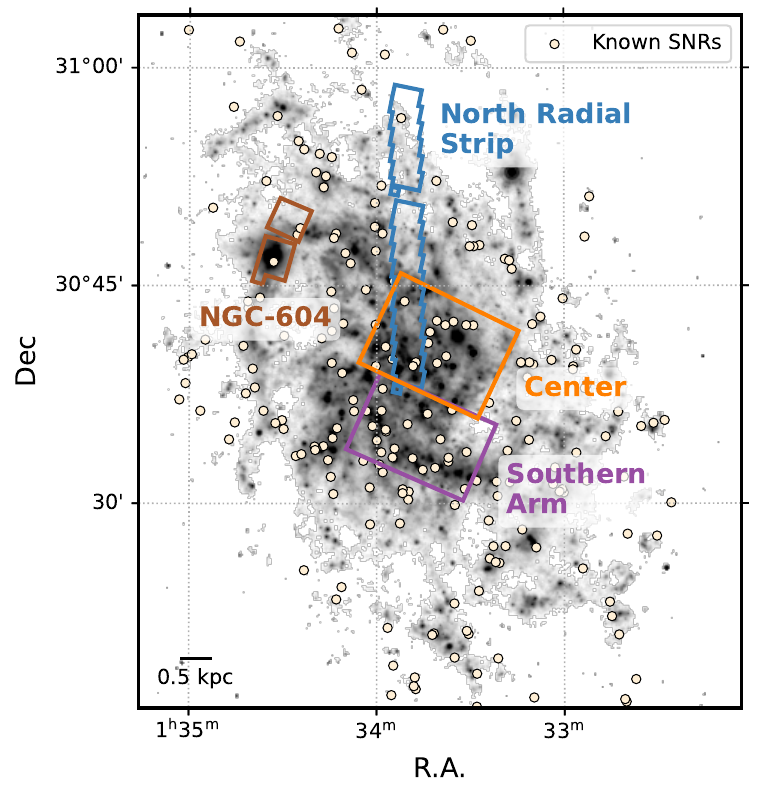}
    \caption{Footprints (colored regions) of the Cycle 1+2 JWST NIRCam and MIRI surveys of M33 that we use for analyzing known SNRs (off-white circles) in this paper. See Section \ref{sec:obs} for details. Greyscale denotes a \Spit 24\mum image of M33.}
    \label{fig:surveyfootprints}
\end{figure}

In this paper, we push the boundary of extragalactic SNR studies with JWST's unprecedented leap in sensitivity and resolution with a new MIRI/NIRCam survey of M33, our closest actively star-forming and gas-rich dwarf spiral galaxy. JWST's potential for new discoveries in SNR science has been demonstrated with recent observations of SN 1987A \citep[e.g.][]{Larsson2023, Jones2023, Arendt2023}, Cas A \citep{Milisavljevic24}, and Crab Nebula \citep{Temim2024}.

Here we report new JWST observations of SNRs in the inner part of M33, showcasing the ability of JWST to detect and resolve IR emission from SNRs beyond the Magellanic Clouds (see Figure \ref{fig:jwst-spitzer-comparison} for a comparison of SNRs with \Spit), allowing access to the substantial SNR population in our Local Group \citep{Long2017}.
Although the JWST wavelengths do not probe the total dust mass as directly as longer wavelength mid- and far-IR emission, with JWST we can carry out detailed studies of shock interaction between the SNR and the dense, dust-rich ambient ISM. A wide variety of lines and continuum features appear in the $\sim$1-25 \mum spectra of interacting SNRs \citep{Reach2006, Williams2006}. This paper presents a ``first-look'' at SNRs in four discrete regions of M33 using a heterogeneous set of NIRCam and MIRI imaging filters drawn from early-mission JWST programs. We focus on what the SNRs look like in the near- and mid-IR and how their properties correlate with known multiwavelength properties.



\begin{figure*}
    \includegraphics[width=\textwidth]{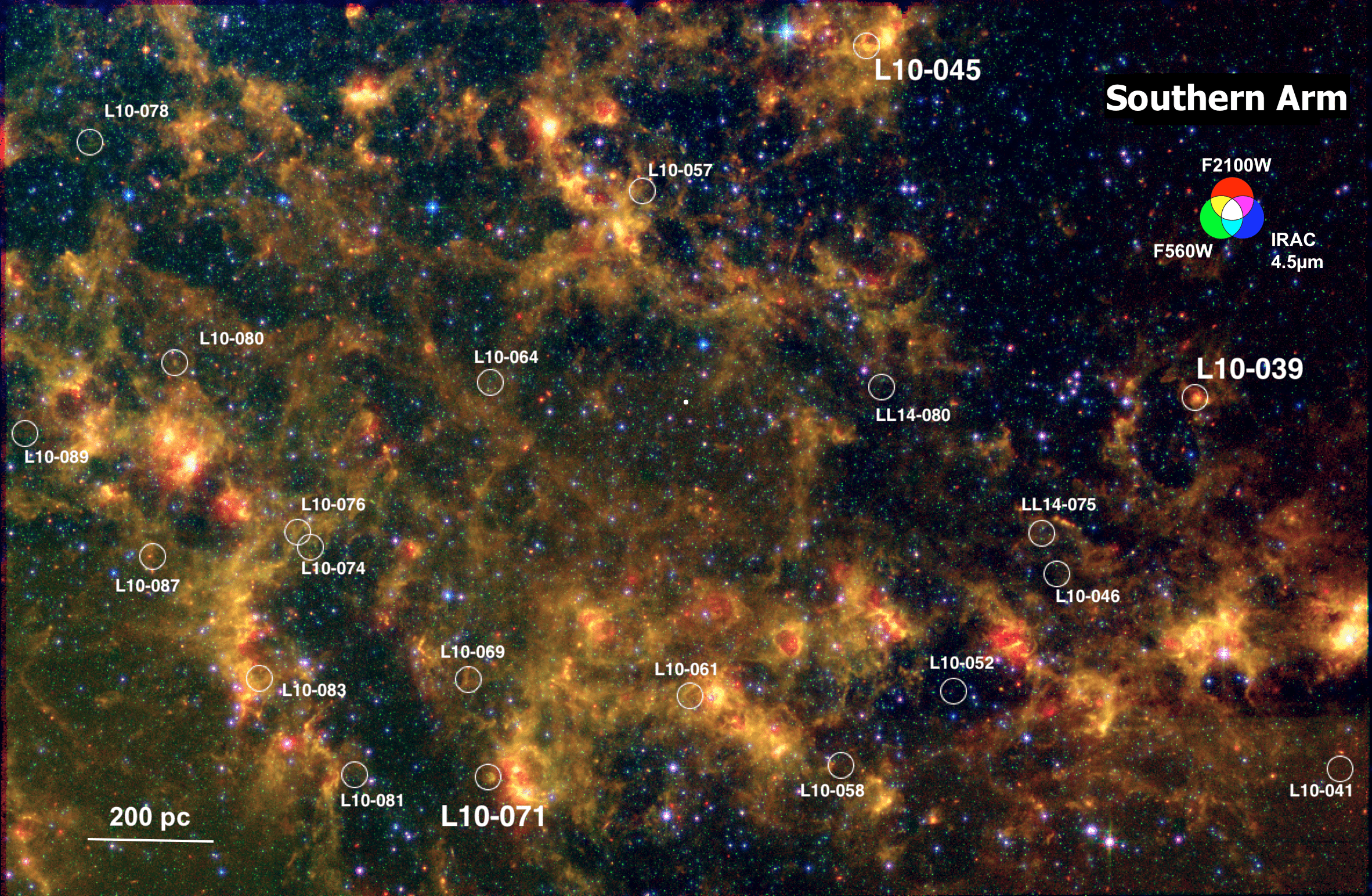}
    \caption{Three-color composite image of the \textbf{Southern Arm} field (Figure~\ref{fig:surveyfootprints}) in our JWST survey of M33, with MIRI filters F560W (green) and F2100W (red), and the IRAC 4.5 \mum (blue). White circles represent locations of confirmed SNRs. The brightest and most prominent MIRI SNRs (see Section \ref{sec:miri}) in the field are labeled in larger font. }      
    \label{fig:miricutout}
\end{figure*}
 \begin{figure*}
    \centering  
    \includegraphics[width=\textwidth]{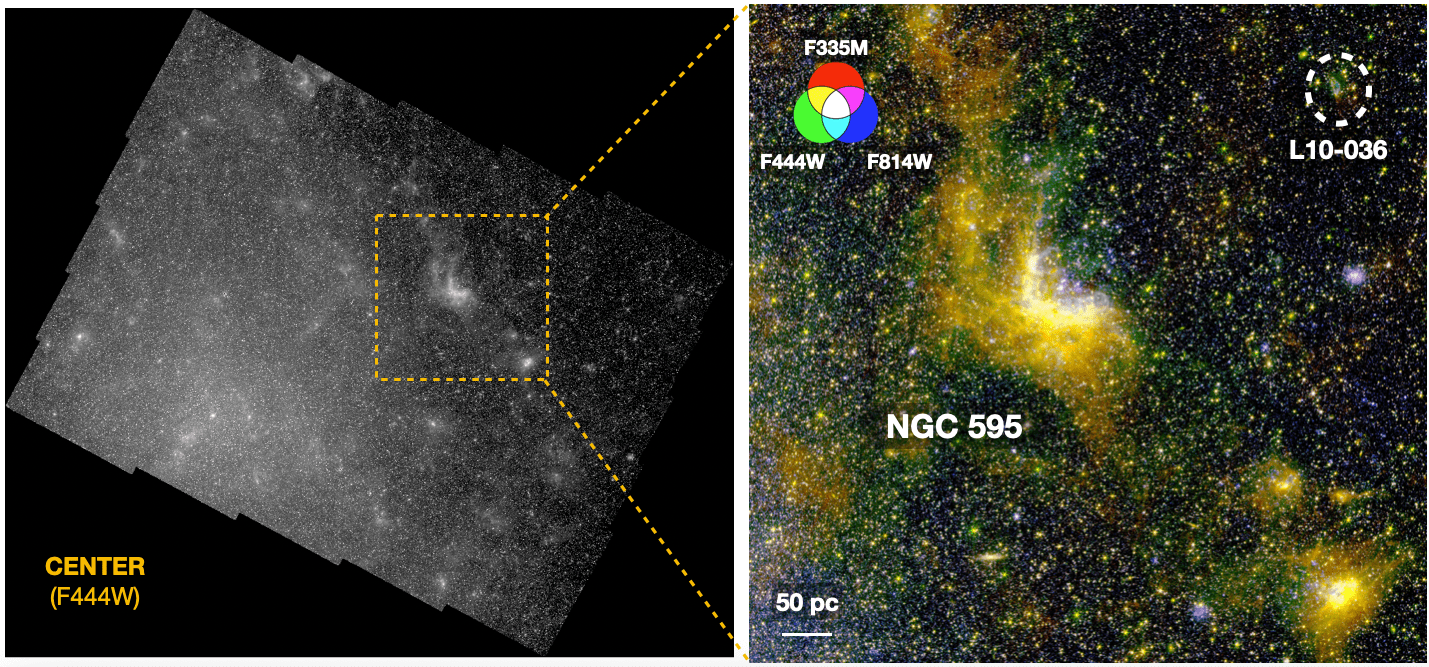}
    \caption{View of the Center field in the NIRCam F444W filter(left) with a zoom-in of the prominent NGC 595 HII region (yellow dashed box) shown on the right. The RGB image was made with the NIRCam F335M (red) and F444W (green) filters, and the Hubble F814W filter from PHATTER (blue). The prominent HII region NGC 595 and our only detected NIRCam SNR L10-036 are labelled. } 
    \label{fig:nircamcutout}
\end{figure*}
Our paper is organized as follows: In Section \ref{sec:obs}, we describe NIRCam and MIRI observations of four discrete fields of M33, along with the multiwavelength datasets we will use to characterize these SNRs. Section \ref{sec:results} describes the observed JWST morphologies of the SNRs in the four fields, their multi-wavelength properties, and any correlation with their progenitor properties. Section \ref{sec:discussion} discusses these observations in the broader context of infrared studies of SNRs and dense gas interactions.

\section{Observations and Methodology} \label{sec:obs}
Our strategy in this paper is to leverage archival JWST observations that happen to overlap with SNRs previously identified in high-quality multi-wavelength surveys. In this section, we describe the JWST observations and how they were reduced, as well as the supporting multi-wavelength observations and datasets that we use to characterize the SNRs. Throughout the paper, we adopt a distance to M33 of 840 kpc \citep{Freedman2001}.

\subsection{JWST data} \label{sec:obs:jwst}
As shown in Figure \ref{fig:surveyfootprints}, we use archival Cycle 1 and 2 JWST datasets covering several regions of M33 in MIRI and NIRCam bands. We will henceforth refer to these as: the MIRI-field, NIRCam-field, the North Radial Strip, and the NGC-604 fields. Basic survey details of these fields, including proposal IDs and observation dates are summarized in Table \ref{tab:survey}, and details of the filters themselves, along with the most common emission sources expected within the passbands are summarized in Table \ref{tab:bands}. Below we provide further description of the spatial and filter coverage in these fields.
\begin{enumerate}
\item \textbf{Southern Arm}: The field was observed with the F560W and F2100W filters, covering a roughly 0.095$\times$0.145 deg$^2$ ($\sim$1.39 $\times$ 2.13 kpc$^2$) patch of the southern spiral arm of M33. The field was constructed from a 5$\times$5 dithered mosaic obtained as part of the GO 2128 program (PI: E. Rosolowsky). The observations were designed primarily with the aim of detecting embedded young massive stars and ISM dust \citep[see][]{Peltonen2024}, although as we demonstrate later, the field contains a number of known SNRs in the region with unprecedented spatial detail in the near/mid-IR.  Both F560W, centered at 5.6 \mum , and F2100W, centered at 21 \mum , are broadband filters that primarily capture  emission from warm, stochastically-heated small dust grains in the interstellar medium. The full-width at half-maxima (FWHM) of the point spread function (PSF) are 0.21\arcsec ($\sim$0.84 pc, for F560W) and 0.67\arcsec ($\sim$2.75 pc for F2100W) at the distance of M33. A detailed view of the full MIRI field, and locations of the known SNRs (see Section \ref{obs:snrcatalog}), are shown in Figure \ref{fig:miricutout}. 

\begin{deluxetable*}{lllll} \label{tab:survey}
\tablecaption{Fields observed with JWST in this paper. Locations of the fields are in Figure \ref{fig:surveyfootprints}. From left to right, columns indicate the field name, JWST instrument, filters observed in the field, proposal ID and observation dates.}
\tablehead{\colhead{Field} & \colhead{Instrument} & \colhead{Filters} & \colhead{Proposal ID} & \colhead{Obs Dates}}
\startdata
Southern Arm & MIRI & F560W, F2100W & GO 2128 & Aug 18, 2022 \\
Center & NIRCam & F335M, F444W & GO 2128 & Aug 18, 2022 \\
North Radial Strip & MIRI & F770W & GO 3436 & Jan 17, 2024 \\
NGC 604 & NIRCam & F090W, F187N, F200W, F335M, F444W, F470N & DD 6555 & Jan 28, 2024 \\
-- & MIRI & F770W, F1130W, F1500W & -- & Jan 26, 2024 \\
\enddata
\end{deluxetable*}

\begin{deluxetable*}{lrrrrrrl} \label{tab:bands}
\tablecaption{JWST filters used in our survey. From left to right, the columns indicate filter name, pivot wavelength ($\lambda_p$), wavelength range ($\Delta \lambda$), fields from Table \ref{tab:survey} where filters were exposed and the corresponding exposure time in seconds, angular resolution in arcseconds ($\Delta \theta$), physical resolution in pc ($\Delta x$) and the major sources of emission expected in these filters}
\tablehead{\colhead{Filter$^e$} & \colhead{$\lambda_p$$^{a}$} & \colhead{$\Delta \lambda$$^{a}$} & \colhead{Field$^e$} & \colhead{Exposure} & \colhead{$\Delta \theta$$^{a,b}$} & \colhead{$\Delta x$$^{c}$} & \colhead{Expected major sources of emission$^{d}$}\\
\colhead{} & \colhead{($\mu$m)} & \colhead{($\mu$m)} & \colhead{} & \colhead{(s)} & \colhead{($\arcsec$)} & \colhead{(pc)} & \colhead{}}
\startdata
F090W & 0.9 & 0.79$-$1.01 & 3 & 2362 & 0.03 & 0.13 & Continuum (stars), Pa$\infty$, [\ion{S}{3}]:0.9,0.95\mum  \\
F187N & 1.87 & 1.86$-$1.88 & 3 & 4724 & 0.06 & 0.26 & Pa$\alpha$ \\
F200W & 1.99 & 1.76$-$2.23 & 3 & 2362 & 0.07 & 0.27 & Continuum (stars), Pa$\alpha$, H$_2$ \\
F335M & 3.36 & 3.18$-$3.54 & 2 & 3221 & 0.11 & 0.45 & PAH, Continuum (stars) \\
... & ... & ... & 3 & 2362 & ... & ... & ... \\
F444W & 4.4 & 3.88$-$4.98 & 2 & 7516 & 0.15 & 0.59 & Br$\alpha$, [\ion{Ar}{6}]:4.53\mum, [\ion{Fe}{2}]:4.65\mum, H$_2$,\\
& & & & & & & Continuum (stars, dust) \\
... & ... & ... & 3 & 2362 & ... & ... & ... \\
F470N & 4.71 & 4.68$-$4.73 & 3 & 4724 & 0.16 & 0.65 & H$_2$ 0-0 S(9), CO \\
F560W & 5.64 & 5.05$-$6.17 & 1 & 14153 & 0.21 & 0.84 & Continuum (stars, dust), H$_2$, PAH \\
F770W & 7.64 & 6.58$-$8.69 & 3 & 699 & 0.27 & 1.10 & PAH, H$_2$, [\ion{Ar}{3}]:6.98\mum, [\ion{Ne}{6}]:7.65\mum \\
... & ... & ... & 4 & 10190 & ... & ... & ... \\
F1130W & 11.31 & 10.95$-$11.67 & 3 & 699 & 0.38 & 1.53 & PAH \\
F1500W & 15.06 & 13.53$-$16.64 & 3 & 699 & 0.49 & 1.99 & PAH \\
F2100W & 20.79 & 18.48$-$23.16 & 1 & 14153 & 0.67 & 2.75 & Continuum (dust), [\ion{S}{3}]:18.71\mum
\enddata
\tablecomments{$^{a}$Based on JWST User Documentation.\\
$^b$Empirical full-width at half-maximum.\\
$^c$Assuming a distance of 840 kpc to M33 \citep{Freedman2001}.\\
$^{d}$Based on \cite{Oliva1999, Reach2006, Neufeld2007, Allen2008, Rho2024}.\\
$^e$(1)--Southern Arm, (2)--Center, (3)--NGC 604, (4)--North Radial Strip}
\end{deluxetable*}

\item \textbf{Center:} This F335M+F444W mosaic was obtained in parallel with the Southern Arm, also as part of GO 2128 (Figure \ref{fig:nircamcutout}). The field covers an area of roughly 0.11$\times$0.15 deg$^2$ ($\sim$$1.6 \times 2.2$ kpc$^2$), at a resolution of 0.11\arcsec ($\sim$0.45 pc) with F335M, and 0.14\arcsec ($\sim$0.57 pc) with F444W. Both filters are sensitive to hot dust emission and spectral lines (e.g. Br-$\alpha$ 4.05 \mum, H$_2$ 0-0) in the ISM, while emission in the F335M filter is sensitive to stellar continuum as well as the 3.3 \mum PAH emission feature arising from aromatic C$-$H stretching modes.
\item \textbf{NGC 604}: We include imaging of this active HII region as it provides the most panchromatic view of IR-visible SNRs in our survey, across multiple NIRCam and MIRI filters. These include: F090W, F187N (tracing Pa-$\alpha$), F200W, F335M and  F444W (same as the Center field), F470N (which sits on the H$_2$ S(9) line and part of the CO fundamental band), and MIRI F770W, F1130W, F1500W, tracing various PAH emission features. The observations were obtained from the public program GO 6555 (PI: M.G. Marin). Two fields were observed: one centered on the HII region NGC 604 observed with MIRI and NIRCam, and an off field to the north observed only with NIRCam. The sky footprints of the NIRCam and MIRI observations of NGC 604 are slightly different, but on average, for both fields, they cover about (0.04$^{\circ}$)$^2$ area, or roughly (0.59 kpc)$^2$, as seen in Figure \ref{fig:surveyfootprints}. The observations provide highly resolved views of the SNR and its surrounding ISM, with spatial resolution ranging from 0.03\arcsec ($\sim$0.12 pc) for F090W to 0.48\arcsec ($\sim$1.9 pc) for F1500W.

\item \textbf{North Radial Strip:} We also include observations of a radial strip of M33 with the F770W filter, which we obtained as part of GO 3436 (PI: E. Rosolowsky). The strip is about 4 $\times$ scale length from the center, spanning a physical area of about 5.25 kpc $\times$ 0.62 kpc. The final map was constructed from a combination of 18 mosaic tiles, and has a spatial resolution (full-width half maximum) of 0.269\arcsec, or roughly 1.1 pc at the distance of M33. Similar to F335M, the F770W filter is dominated by the 7.7 \mum PAH feature, which arises from a combination of C-C stretching and C-H in-plane bending modes. The mosaic design was constructed with the purpose of tracing the spatial and radial variation of diffuse PAH emission on pc-scales, although we will mainly focus on the few SNRs that coincide with the strip. 
\end{enumerate}
 
Observations were reduced using the \texttt{pjpipe} reduction pipeline designed for the Physics at High Angular Resolution in Nearby GalaxieS (PHANGS)-JWST survey \citep[see][for details]{Lee2023, Williams2024, Peltonen2024}. The \texttt{pjpipe} package uses the official JWST pipeline \citep{jwst-pipeline} as a base, but includes some changes to the default pipeline parameters and the introduction of several custom steps that optimize the imaging of large fields filled with extended emission. Compared to the official pipeline, \texttt{pjpipe} notably implements different background matching approaches, a multistep destriping algorithm for NIRCam data, and a tiered set of astrometric alignment choices to produce good alignment between different JWST images of the same region.  We use Gaia DR3 \citep{gaia_dr3} sources to set the absolute astrometric alignment for all bands, and we find there are sufficient numbers of Gaia sources that have mid-IR counterparts to get good ($\lesssim$0.05\arcsec) global alignment. We estimate this degree of alignment by measuring the relative positions of stars in the F770W compared to the Gaia-aligned HST data from \citet{Williams2021}.

Where available (programs 2128, 2130, 3436), we use MIRI background observations from the same programs to correct the observations for the variable mid-IR background.
While \texttt{pjpipe} implements a flux anchoring algorithm to set the flux calibration zero-flux level on the resulting images, this is primarily relevant for characterizing extended emission. As we do not carry out analyses that rely on an absolute flux from the JWST data, we do not carry out the anchoring step for the images shown here, so there can still be small zero-point offsets from optimal flux calibration \citep{Williams2024}.  

\subsection{Supernova Remnant catalog} \label{obs:snrcatalog}
We specifically focus on the JWST observations at the locations of known SNRs in M33. These SNRs have been assembled from decades of multi-wavelength observations \citep{Dodorico1982, Long1990, Gordon1999, Long2010, Lee2014, Garofali2017, Long2018, White2019}. The SNRs were primarily identified from line ratios of \siiha from narrowband images, and further confirmed with optical spectroscopy, radio  and X-ray data. The most up-to-date compilation of SNRs in M33 appears in \cite{White2019}, which adds high sensitivity Very Large Array (VLA) radio observations at 1.4 and 5 GHz to the optical and X-ray identified SNRs from \cite{Long2018} and \cite{Long2010}. We use the list of SNRs from \cite{White2019} to cross-reference with our JWST fields. At various points in this paper (particularly in Section \ref{sec:res:multiwavelength}), we will use published measurements from these papers to characterize the JWST-visible SNRs. We will particularly rely on measured fluxes and ratios of optical emission lines from \cite{Long2018}, which compiles new multi-fiber spectroscopy from HectoSpec from MMT for each SNR, along with archival spectroscopy from \cite{BK85, Smith1993, Gordon1998}, and \cite{Long2010} for some SNRs. We refer the reader to these papers for details of the individual SNR observations, and to \cite{Long2017} for the historical progression in the M33 SNR catalog construction. 


\subsection{Supporting Multi-wavelength Images} \label{sec:obs:multiwavelength}
In order to identify the extent of the SNRs and their surrounding ISM and stellar populations in the JWST images, we will make use of several published (and in-preparation) multi-wavelength images of M33.
\subsubsection{Ground-based Narrowband Images}
We use published optical images as reference for the SNR morphology in the JWST images. Radiative shocks in SNRs with speeds of $\sim$50-500 km/s produce bright forbidden line emission of various elements from near-UV to mid-IR wavelengths \citep[e.g.][]{Raymond1979, Dopita1995, Allen2008}, and these are useful for identifying and tracing the structure of SNRs \citep[e.g.][]{Mat73, Fesen1985, Mat97}. 
Our main reference images are narrowband continuum-subtracted images of \ha, \sii and \oiii from the Local Group Galaxies Survey (LGGS), obtained with the 4-m Mayall telescope with typical seeing-limited spatial resolution of $\approx$1\arcsec \citep{Massey2006, Massey2007, Massey2007b}, or roughly 4 pc at the distance of M33. While radio and X-ray images can also be used for reference, only a small subset of the M33 SNRs are visible in X-ray, and many of the largest, low optical-surface brightness SNRs are not detected in radio \citep{White2019}. Finally, the narrowband LGGS images are the highest resolution images of emission line nebulae available in M33 at the time of writing this paper. As a result, we mainly rely on optical images as a reference for SNR morphologies. We use continuum-subtracted images of \ha and \sii using a scaled $R$-band image, and continuum-subtracted \oiii using a scaled $V$-band image from LGGS.

\subsubsection{Additional Radio/mm data}
While the images and datasets described above provide adequate supporting information for our SNRs, we show high-resolution archival mm data (e.g., CO(2-1), CO(1-0)) and radio 1.4 GHz data for select JWST-detectable SNRs that we felt could benefit from additional information. The CO data reveal the dense, molecular ISM distribution around the SNRs, while the 1.4 GHz radio images can reveal the extent of the SNR blast-wave if the optical \ha images appear insufficient.

For SNR L10-036, we use archival $0\farcs5$ (2~pc) ALMA CO(2-1) data from project 2018.1.00378.S, previously published in \citet{Sano2021}. We reprocessed these data from the ALMA archive and used the PHANGS-ALMA imaging pipeline to make the products presented here \citep[see][]{Leroy2021_pipeline}. The images shown in this work use CASA's \texttt{tclean} to image and deconvolve the visibilities, with Brigg's weighting with \texttt{robust=0.5} in $1.2$~km s$^{-1}$ channels. The CO data cube has an rms of 38 mK per $1.2$~km s$^{-1}$ channel. We create an integrated intensity mask following the PHANGS-ALMA prescription, specifically the ``strict mask'' criteria.

For the NGC 604 SNRs L10-118 and LL14-168, we also show ALMA CO(1-0) data with a spatial resolution of $\sim$4\arcsec ($\sim$16 pc) to characterize the ISM near the blast-waves. These data are from ALMA project 2022.1.00276.S, which surveyed the PHATTER region (section \ref{sec:obs:hstsfh}) using the 12m+7m+total power arrays.  The data were calibrated using the observatory pipeline (v.\ 6.5.4.9) and imaged using the PHANGS-ALMA pipeline \citep{Leroy2021_pipeline} using the same parameter choices as for the CO(2-1) data. Both calibration and imaging used \textsc{CASA} \citep{CASA}.  The resulting data have a synthesized beam FWHM of $3.9''$, a velocity channel width of 1.3~km s$^{-1}$, and a typical noise value of 0.3~K in each channel. Again, we create an integrated intensity map for the CO(1-0) using the strict masking.

We also show new radio continuum 1.4 GHz data taken by the Very Large Array (VLA) for these two SNRs from the Local Group L-Band Survey (LGLBS). Details of the survey will be published in the upcoming survey description papers \citep[][Koch+ in prep, Sarbadhicary+ in prep]{Koch2024}, but we provide some relevant details here. LGLBS (VLA project 20A-346, PI: A. Leroy) has collected 2200 hrs of 1.4 GHz observations in the A, B, C, and D configurations for six Local Group galaxies (including $\sim400$~h of archival VLA observations). 
M33 was observed in 13 L-band pointings (1-2 GHz with 16 spectral windows of 64 MHz), with each pointing having a total of 38.5 hrs in all the configurations. Continuum data were reduced, calibrated, quality-assured, and split with a customized workflow based on the standard VLA CASA pipeline. We separately imaged the B+C+D data in each field with \texttt{tclean} using wide-band (\texttt{deconvolver=mtmfs, nterms=2}), and wide-field (\texttt{gridder=wproject, wprojplanes=512}) imaging algorithms out to 1\% of the primary beam. The Briggs-weighted (with \texttt{robust=0}) total-intensity images of each field were cleaned to a 1$\sigma$ depth of $\sim$4 $\mu$Jy, with restored beams of 4-5\arcsec. The images were then linearly mosaicked with a Sault-weighting scheme\footnote{\href{https://library.nrao.edu/public/memos/vla/sci/VLAS_154.pdf}{Sault (1984)}} using the 1.4 GHz primary beam patterns. These new VLA images exceed the sensitivity and resolution of those from \citet{White2019}\footnote{\url{https://sundog.stsci.edu/m33/}}.
In particular, the LGLBS imaging is able to capture fainter, more spatially-extended structures as a result of the combined baselines from B+C+D configurations with excellent uv-coverage in each field.

\subsection{HST Images and Star-formation histories} \label{sec:obs:hstsfh}
\begin{figure*}
    \includegraphics[width=\textwidth]{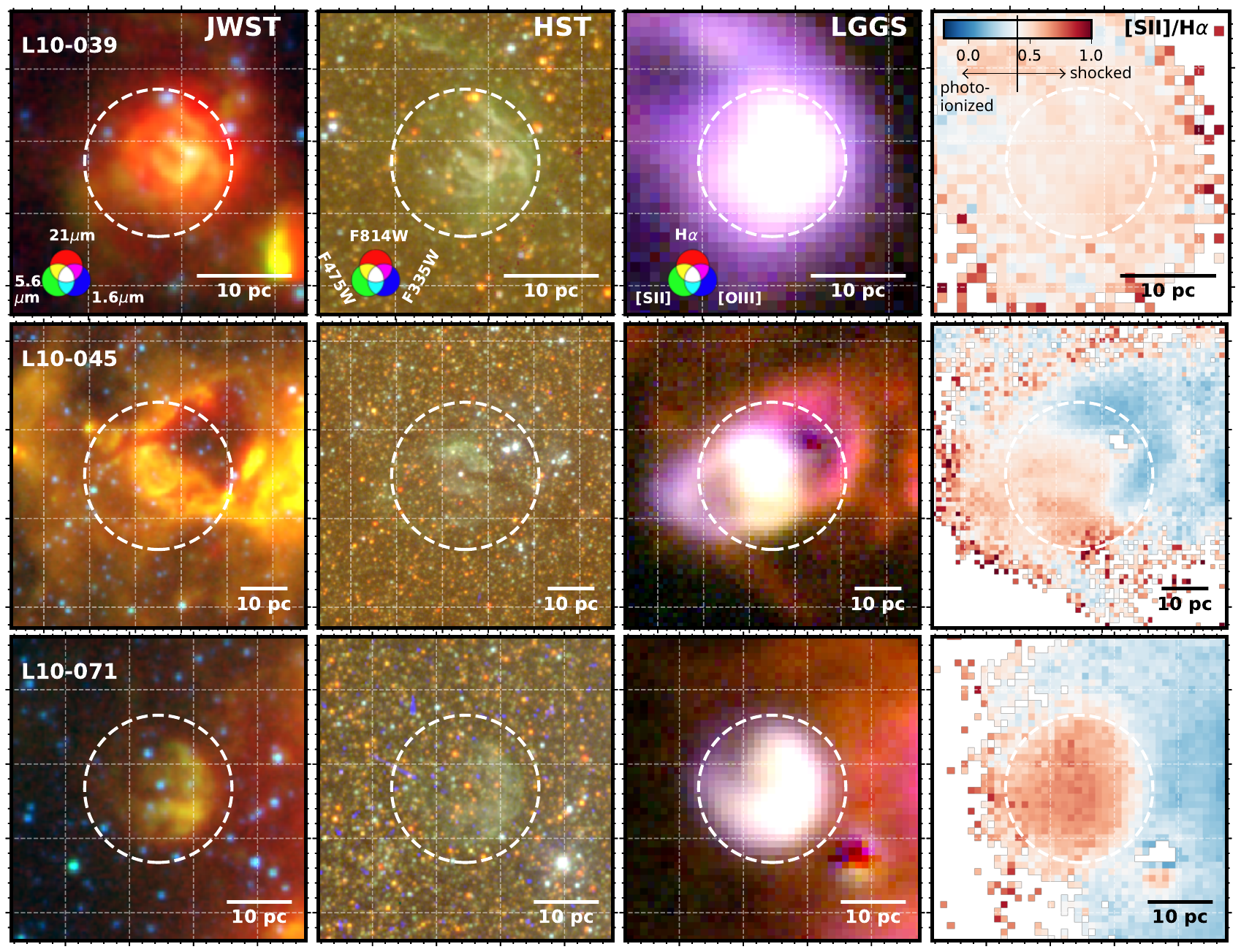}
    \caption{Cutouts of the most prominent SNRs in the Southern Arm. \emph{Left}-to-\emph{Right}: three-color image using HST F160W, JWST F560W and F2100W, three-color HST image using the F335W, F475W and F814W filters, three-color image using narrowband continuum-subtracted \oiii, \sii and \ha images, and an \sii/\ha ratio image with the colorscale centered on \sii/\ha=0.4, which is the dividing value (vertical line with arrows) assumed here between photoionized regions (bluer colors) and shock-heated regions (redder pixels). The Venn diagrams denote the filter combinations used to produce the RGB images. Dashed circles sketch the diameter of the SNR reported by \cite{White2019} as estimated in lower-resolution surveys (Section \ref{obs:snrcatalog})}
    \label{fig:brightmirisnrs}
\end{figure*}

For the brightest SNRs, we also make use of the much higher-resolution wide-band HST images from the PHAT Triangulum Extended Region (PHATTER) survey \citep[][GO 12055, 14610; PI: J. Dalcanton]{Williams2021}, specifically the WFC3 UVIS F336W, ACS WFC F475W and F814W filters, and WFC3 IR F160W filters with a typical full-width at half-maximum of $\approx$0.07\arcsec ($\sim$0.3 pc at M33)\footnote{\url{https://hst-docs.stsci.edu/wfc3ihb/chapter-6-uvis-imaging-with-wfc3/6-6-uvis-optical-performance}}. Although SNRs are not broadband emitters at optical wavelengths, their strong emission lines can be picked up in the wide-band filters as we demonstrate later in Figure \ref{fig:brightmirisnrs}. The high angular resolution HST images of these SNRs allow a more accurate structural comparison with JWST than the lower-resolution LGGS images. The relevant spectral lines from shocks that appear within the above filter bandpasses are nicely illustrated in Figure 3 of \cite{Lin2020}, based on predictions from the MAPPINGS III shock library of \cite{Allen2008}. We expect the F336W filter to trace [\ion{Ne}{5}]$\lambda$3426, F475W to trace \oiii $\lambda$4959,5007 and \hb $\lambda$4861, also possibly some [\ion{He}{2}]$\lambda$4686 and [\ion{N}{1}]$\lambda$5200, and F814W to trace lines from the Pa-series of H, and the [\ion{S}{3}] $\lambda$9069, 9532 lines. The strengths of these lines strongly depend on  shock velocity and ambient density, and less strongly on factors like the magnetic field, composition and ionization states of the ambient material, and atomic physics \citep{Allen2008, Sutherland2017}.

We also make use of the PHATTER star-formation histories \citep[SFHs,][]{Lazz2022} to check how the visibility of the SNRs in the JWST images could be related to the recent SFH ($<$56 Myr) in their vicinity. In this paper, we will be discussing these age ranges in terms of their equivalent stellar ``mass'' range, as measured by \cite{Koplitz2023} (hereafter, K23) from the PHATTER SFHs. These were interpreted by K23 as the most probable progenitor masses of the SNRs. We use the median and 68th percentile progenitor mass range of each SNR reported in Table 2 of K23, derived from conversion of the mass-weighted stellar age range in the SNR vicinity assuming single stellar evolution models. We refer the reader to K23 for the details of these measurements. Note that K23 reports two types of progenitor masses for each SNR. The first set of masses, which we refer to as ``grid-based'', are based on the location of the SNR in the \cite{Lazz2022} SFH map, where the SFHs are calculated in a grid of 100$\times$100 pc cells. The second one, which we refer to as ``best-fit'', are from more localized SFHs measured by K23 using only stars within 50 pc of each SNR, after subtracting a `background' SFH (from a 50pc to 1 kpc annulus centered on the SNR). The latter accounts for contamination by background stellar populations that are likely unrelated to the SNR. The lowest measurable progenitor mass, by definition, will be 7.1 \Msun (corresponding to 56 Myr). Some SNRs have no best-fit progenitor mass above this mass-range, indicating that there is no recent star formation uniquely associated with the SNR. We will show these SNRs as having best-fit upper limit masses of 7.1 \Msun. 

Note that the true probability of old stellar progenitors of these SNRs is unconstrained as the PHATTER SFHs have only been measured up to 0.6 Gyr \citep{Lazz2022}. The recent ($<$50 Myr) SFH however is more relevant to our analysis, as the correlation between SN explosions and ISM sites, which as we show later is the main driver of the infrared emission, becomes more and more randomized for lower mass stars \citep{Sarbadhicary2023}.

\begin{figure*}
    \centering
    \includegraphics[width=0.9\textwidth]{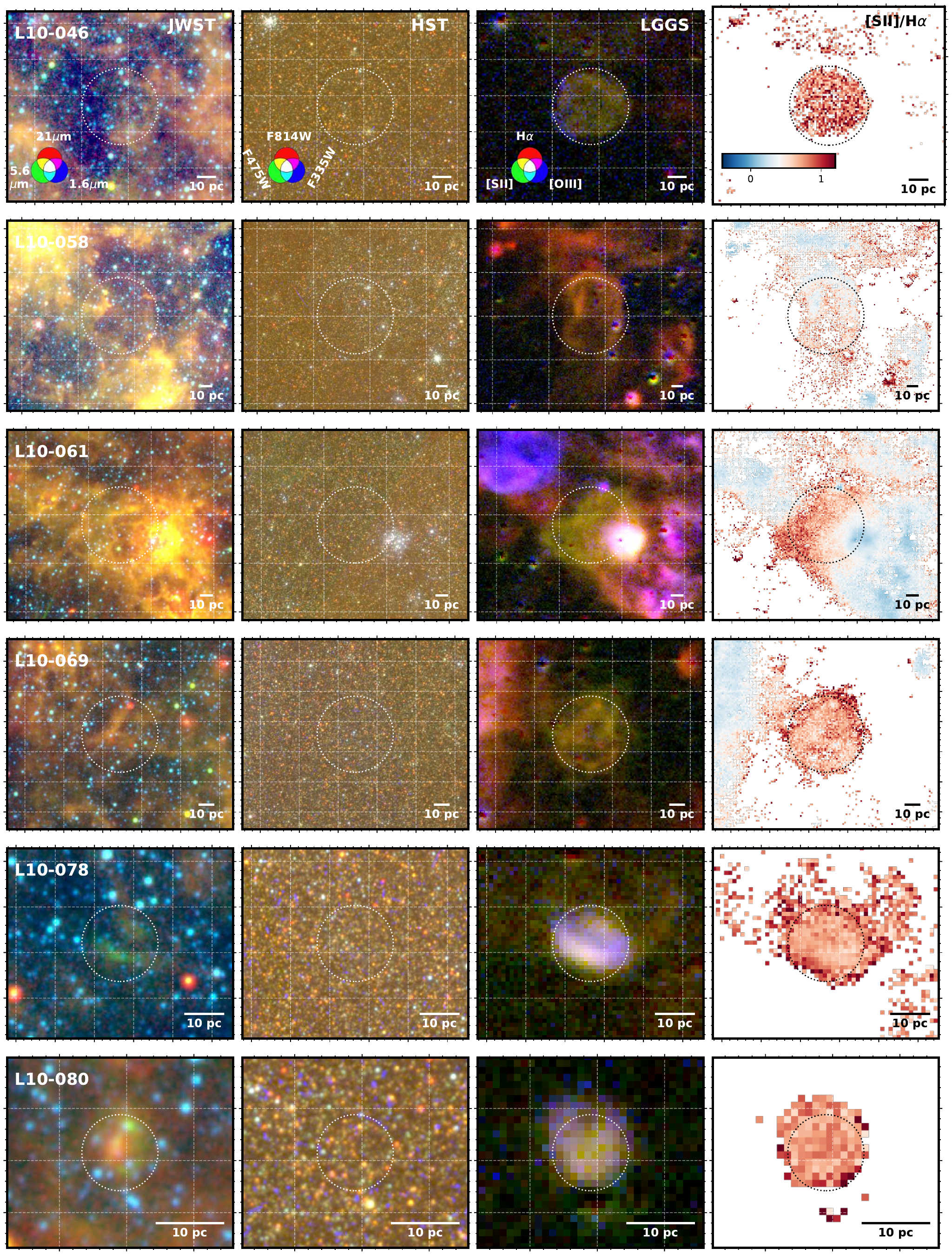}
    \caption{Partially detected MIRI SNRs. Images are the same as Figure \ref{fig:brightmirisnrs}, with three-color JWST, HST, LGS and \siiha ratio.}
    \label{fig:partialsnrs}
\end{figure*}
\begin{figure*}
    \centering
    \includegraphics[width=\textwidth]{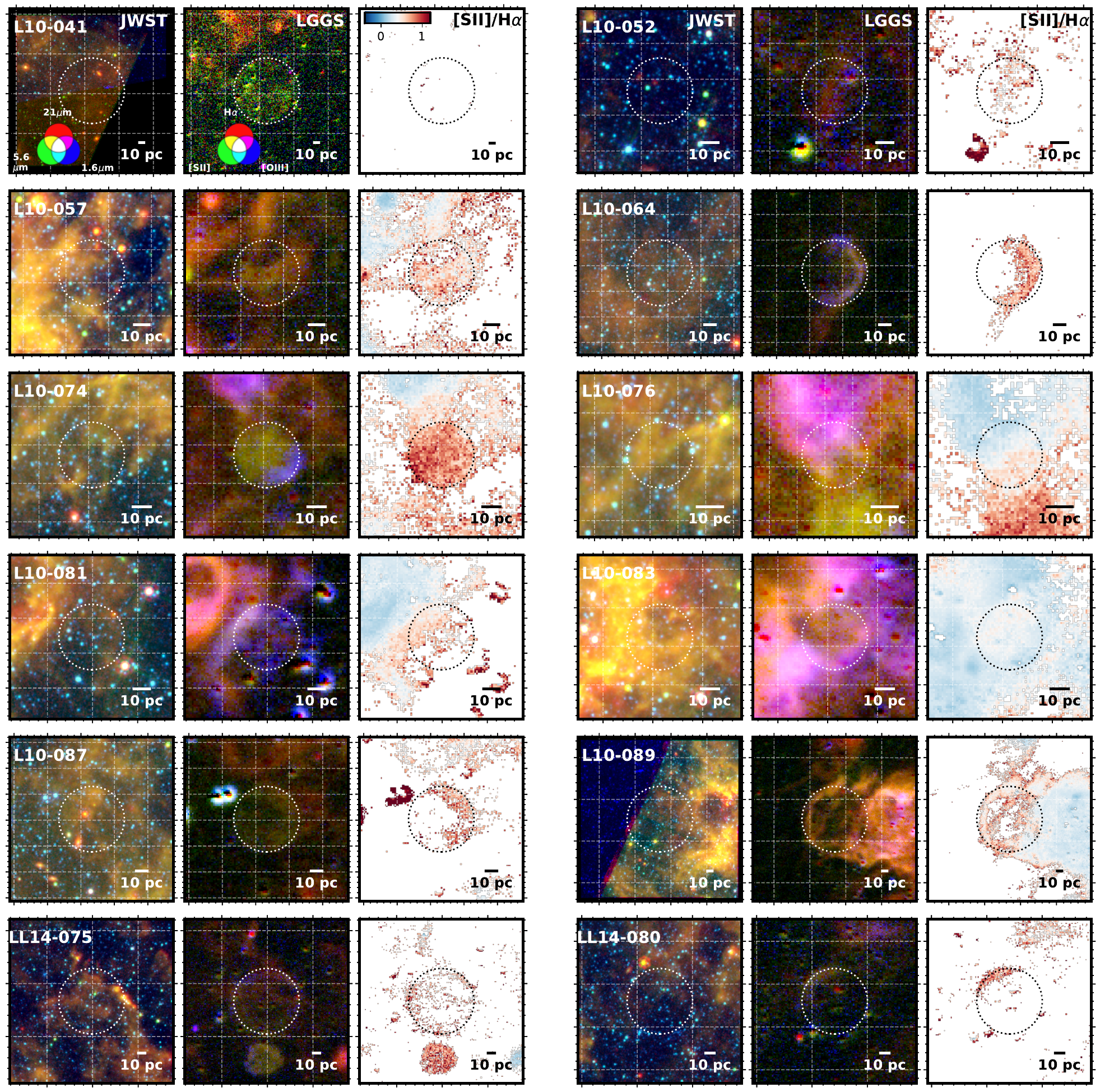}
    \caption{Ambiguous/non-detection SNRs in the Southern Arm field of M33. Similar to Figure \ref{fig:brightmirisnrs} and \ref{fig:partialsnrs}, we show three color JWST, LGS and \siiha, but omit the HST images since none of the SNRs show any visible emission.}
    \label{fig:nondetectionmirisnrs}
\end{figure*}

\subsection{Methodology} \label{sec:obs:method}

In this paper, we mainly study the visibility and spatially-resolved structures in the JWST images, and how their appearances correlate with their known properties from other multi-wavelength surveys. We carefully inspect the SNR sites in the individual JWST filter images by-eye in DS9, adjusting the stretch and data range to look for any emission tracing the known structure of the SNRs. The by-eye approach is necessary as the emission nebulae of the SNRs can often be faint, have indistinct morphologies, and overlap with stars, diffuse ISM filaments, bubbles, and HII regions. More sophisticated identification algorithms for SNRs are in development \citep[e.g.][]{DeBoom2024}, and will be used in future for quantitative studies in the JWST images, but these are presently outside the scope of this paper.

As a reference for the full structure of the SNRs, we use the narrowband \ha and \sii images (and HST images for the brightest SNRs). The \siiha ratio image can help distinguish overlapping regions of photoionized and shock-heated gas, since the former typically has ratios $< 0.4$, while the latter has rations $> 0.4$. A minor caveat is that the LGGS \ha filter is 50 \AA\ wide \citep{Massey2006}, so it may contain a small contribution from the weak [\ion{N}{2}] 6548\AA\ line. Nevertheless, the LGGS narrowband filter ratios do a decent job of distinguishing shock-heated and photoionized nebular regions for our purposes as we show in Section \ref{sec:results}, and have been reliably used before for identifying SNRs in M31 and M33 \citep{Lee2014}. To be certain however, we will occasionally refer to results from published optical spectroscopy on the same SNRs \citep{Long2010, Long2018, SITELLEM33} to confirm the \siiha ratios in cases where the narrowband filter ratios appear ambiguous.

Visual inspection reveals that while some SNRs can be very clearly seen in the JWST images, others are partially seen, or somewhat ambiguous against the background emission. For discussion purposes, we categorize the SNRs into one of four categories (primarily for discussion of the Southern Arm and Center field SNRs),
\begin{itemize}
\item \textbf{Clearly-detected}: the SNR is detected in all the filters in the field, and the detected emission almost perfectly traces out the structure of the SNR in the \ha images
in the regions where \siiha$>$0.4 (e.g.\ L10-039, Figure \ref{fig:brightmirisnrs}).
\item \textbf{Partially-detected}: the SNR is detected in one or more filters, but the detected emission traces only a portion of the \ha structure (e.g.\ L10-069, Figure \ref{fig:partialsnrs}). 
\item \textbf{Ambiguous} (only for Southern Arm field): the SNR has some emission along its line of sight, but it is unclear whether the emission is from the SNR or the background or foreground ISM. This is particularly applicable to the Southern Arm field as the bright dust-rich ISM fills a substantial fraction of our observed area (e.g.\ L10-074, Figure \ref{fig:nondetectionmirisnrs}).
\item \textbf{Non-detections}: No extended emission is visible at the location of the SNR (e.g.\ L10-052, Figure \ref{fig:nondetectionmirisnrs}).
\end{itemize}

We note that this classification is somewhat subjective and user-dependent, and as shown later, dependent on the infrared wavelength range as well. More quantitative or citizen-science based classifications can be performed in future, but for the purposes of first presentation and discussion of images in this paper, we believe the above visual classification is adequate. Similar visual classification schemes have also been used for studying Magellanic Cloud SNRs in infrared images \citep[e.g.][]{Seok2013,Matsuura2022}. 

\section{Results} \label{sec:results}
Here we describe the characteristics of the SNRs in the four JWST fields (Section~\ref{sec:obs}), and assess their correlation with the known multiwavelength properties from published data (Section \ref{obs:snrcatalog}).

\subsection{SNRs in the Southern Arm field} \label{sec:miri}
\begin{figure*}
    \includegraphics[width=\textwidth]{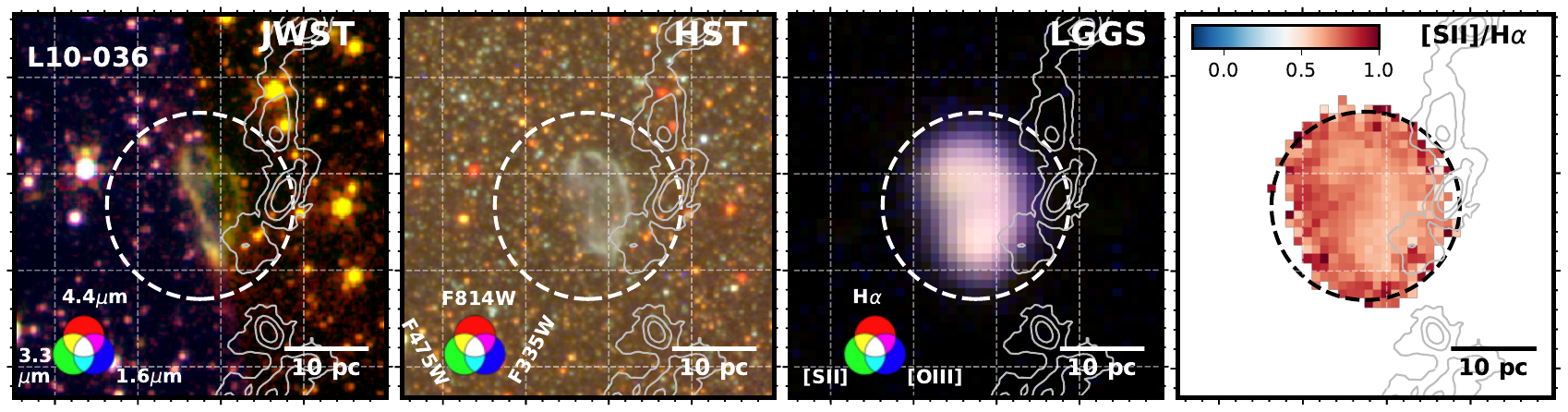}
    \caption{JWST-HST-narrowband LGGS images of the NIRCam-detected SNR L10-036. RGB image scheme follows the same as Figure \ref{fig:brightmirisnrs}, except the JWST MIRI filters are replaced with our NIRCam filters F160W (red), F335M (green) and HST F814W (blue). Emission in these filters are likely from shock-excited ionic and molecular lines summarized in Table \ref{tab:bands}. The gray contours refer to ALMA CO(2-1), described in Section \ref{sec:obs:multiwavelength}. The levels are 3.7, 35.8 and 67.9 \Msunpc (assuming $\alpha_{\mathrm{CO}}=9.74$ M$_{\odot}$ pc$^{-2}$ (K km s$^{-1}$)$^{-1}$ for the metallicity of 0.6 Z$_{\odot}$ in M33).}
    \label{fig:nircamsnr}
\end{figure*}

\begin{figure*}
    \centering
    \includegraphics[width=\textwidth]
    {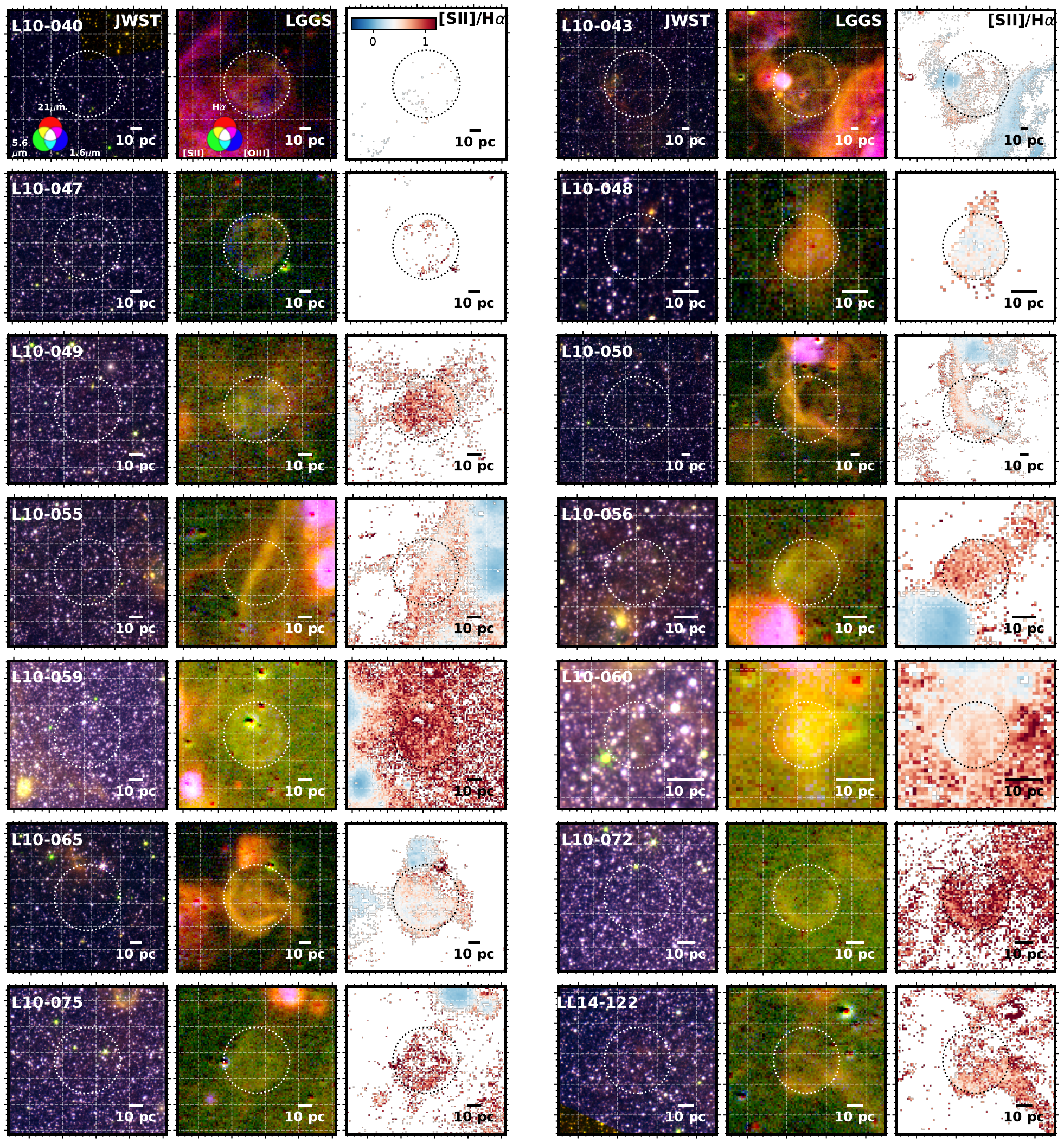}
    \caption{The 14 SNRs other than L10-036 in the Center field. Each SNR has three images -- a three color   
    ``JWST'' near-IR image (HST F160W, NIRCam F335M, and F444W), three color ``LGGS'' \oiii, \ha and \sii, and the \siiha ratio image. As before, the white dashed circles represent the approximate published diameter of the SNR. In the \siiha image, red primarily represents shocked-gas pixels (\siiha$>$0.4) and blue represent photoionized pixels (\siiha$<$0.4).} 
    \label{fig:nondetectionnircamsnrs}
\end{figure*}

Of the 21 SNRs in the Southern Arm field, we find 3 clearly-detected SNRs -- L10-039, L10-045 and L10-071 -- shown in Figure \ref{fig:brightmirisnrs}. The snapshots in Figure \ref{fig:brightmirisnrs} show the RGB images made from: (1) JWST F560W and F2100W filters together with the HST F160W filter from PHATTER, (2) PHATTER HST images from the broadband F335W, F475W and F814W filters and (3) narrowband filter images from LGGS in the lines of \ha, \sii and \oiii\footnote{We include \oiii mainly to complete the RGB images and because photoionized nebulae are richer in atoms at higher ionization states, but the presence of \oiii itself is not a strong diagnostic for distinguishing shocked from photoionized nebulae.}. The SNR morphology is not always easy to distinguish from the surrounding diffuse ISM and HII regions in the emission-line and IR images. To approximately guide the eye, we also display the \sii/\ha ratio image of each SNR. Regions with \sii/\ha$>0.4$ (redder pixels) are expected to be shock-heated, while \sii/\ha$<$0.4 (bluer pixels) are more likely to be photoionized. Note that some pixels with \sii/\ha$>0.4$ are not related to SNRs, and are instead caused by poorly-subtracted stars. Diffuse ionized gas, which has low-surface brightness pixels in \ha, can also show \sii/\ha$>$0.4, and will often be found at the rims of brighter HII regions \citep[e.g.][]{haffner2009warm, blair1997identification, long2022supernova}. We masked all pixels in the \sii and \ha images that were less than 3 times the RMS noise in the image before creating the ratio image.

Figure \ref{fig:brightmirisnrs} shows the close match between the infrared and optical morphologies of the SNRs when comparing the JWST images with the LGGS, and in particular, the broadband HST images. All three SNRs are roughly spherical but with complex morphologies and discontinuities. They are also bright in the narrowband LGGS images, including in \oiii (which is likely what produces the strong greenish-white HST emission, signifying strong radiative shocks interacting with dense gas as we will see later). These SNRs are so bright that their emission sticks out in the broadband filters, even with significant stellar contamination. None of the other SNRs in the field show any such visible emission in broadband HST, as discussed later. All three SNRs are close to bright star-forming regions seen in Figure \ref{fig:miricutout}.

The six partially detected SNRs are shown in Figure \ref{fig:partialsnrs}. For these SNRs, 
the entire SNR is often not visible in JWST, and/or harder to distinguish from the emission of the background ISM, unlike the previous three. The SNRs in this category are a more heterogeneous mix of objects than the clearly-detected cases. For example, L10-069 has a single filament of the \ha morphology with a counterpart in the JWST images, and is likely part of the shock that is interacting with a dense filament of gas along its path. Similarly, L10-078 has a prominent bright filament of F560W in the south that coincides with the luminous southern shock in the LGGS image. The northern part of the shock however has fainter F560W and F2100W emission that appears somewhat blended with the surrounding filaments of ISM. An interesting SNR is L10-080 (last panel, Figure \ref{fig:partialsnrs}), also the smallest SNR in the catalog based on the measured size in \cite{Long2010}. The centrally concentrated 21\mum emission makes it a potential candidate for ejecta dust emission (see further discussion in Section \ref{sec:l10-080}). In other cases like L10-046 and L10-058, parts of the shock are visible but the SNRs are much larger (and sometimes with more disturbed morphologies like the case of L10-058), and embedded in a bright ISM background. 

Further descriptions of the individual clearly-detected and partially-detected SNRs are given in Appendix \ref{app:individualsnrs}.

The ambiguous class of MIRI SNRs are shown in Figure \ref{fig:nondetectionmirisnrs}. None of the SNRs show up in HST images due to their faintness, as was already seen for the partially detected SNRs, so we exclude the HST images from the snapshots. We stress here that `ambiguous' does not mean non-detection of IR emission. While some of the SNRs have no detectable extended emission (e.g. L10-041, LL14-052), many of the others have poorly defined optical morphologies in addition to often being near confusing sources, making identifying any unique mid-IR emission associated with the SNR all the more difficult. One of the SNRs, LL14-075, appears as a large \ha-emitting shell of 90 pc diameter, and encloses some distinct mid-IR emission within its aperture, but this emission does not quite trace the \ha shell and in fact appears interior to the shell, so we have classified it as ambiguous. 

For a few SNRs (e.g.\ L10-057, L10-074, L10-076, L10-089), the optical emitting region of the SNR appears as a gap in the bright IR emission from their surrounding star-forming region. One possibility is that they had simply exploded in the lower density outskirts of the star-forming region with little or no dust \citep[e.g][]{Mayker2023, Sarbadhicary2023}. The other possibility is that the dust was destroyed by the now-evolved blastwave. Indeed evidence of SNRs heating and destroying dust has been inferred in the Magellanic Clouds \citep[e.g][]{Laki2015}, though in our case we cannot quite confirm this with our M33 objects as the spatial resolution of the existing far-IR images of M33 are much larger than the SNRs. 

\subsection{SNRs in the Center field}
The Center field (Figure \ref{fig:nircamcutout}) only has 1 clearly-detected SNR (L10-036, Figure \ref{fig:nircamsnr}) out of the 15 known SNRs in survey area (shown individually in Figure \ref{fig:nondetectionnircamsnrs}). SNR L10-036 is already quite unusual with its elongated, bean-shaped appearance. The shape was noted before in the low-resolution LGGS images \citep[e.g.][]{Long2010} as well as in HST images by \cite{Lin2020}. The NIRCam images reveal that this shape is preserved even in the near-infrared. The shell-like appearance matching the full extent of the shock implies that the IR emission is produced at the forward shock interacting with the ISM. The ubiquitous emission across the optical range, as shown by three-color HST image and narrowband LGGS image in Figure \ref{fig:nircamsnr}, indicates that a signifcant fraction of the shock has become radiative, emitting bright forbidden-lines. The SNR is located about 310 pc (projected) away from the massive HII region NGC 595. We believe the bright IR emission in L10-036 and the unusual shape is most likely arising from interaction with an inhomogenous environment. The SNR is likely already ISM-dominated based on X-ray spectra \citep{Garofali2017}, and appears in projection close to a filament of $^{12}$CO(2-1)-emitting molecular gas west of the SNR (grey contours in Figure \ref{fig:nircamsnr}; the SNR is also optically brighter on this side). 

The remaining SNRs in the Center field do not show any noticeable extended emission that clearly traces the SNR-emitting region like L10-036 (Figure \ref{fig:nondetectionnircamsnrs}). Most of these objects are dominated by stars in the field. Hints of faint emission appear in cases like L10-056 and L10-060 in Figure \ref{fig:nondetectionnircamsnrs} (also see Figure \ref{fig:f770wstrip}), but these emission features could also be unrelated emission from PAH and spectral lines in the surrounding ISM. Some visible emission in F444W is present in some SNRs e.g.\ L10-043, L10-065, but comparison with the optical images show that these are from adjacent HII regions. We conclude therefore that this remaining sample in the Center field are non-detections.
 \begin{figure*}
 \centering
    \includegraphics[width=0.8\textwidth]{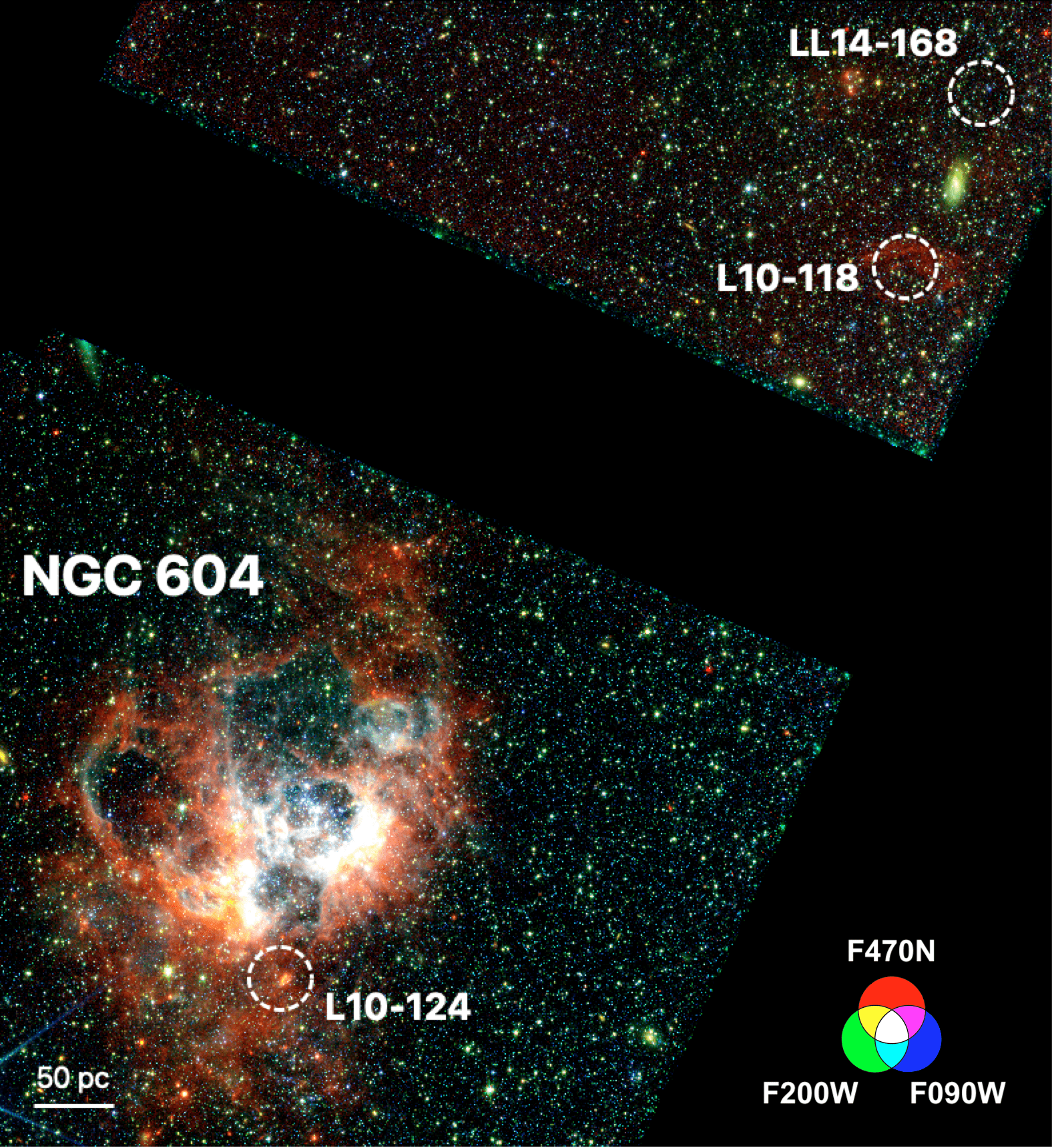}
    \caption{RGB image of JWST observations of NGC 604 in the NIRCam F090W (blue), F200W (green) and F470N (red) filters. Locations of the three SNRs coincident with this mosaic are labeled and circled.}
    \label{fig:rgbngc604}
\end{figure*}

 \begin{figure*}
 \centering
    \includegraphics[width=0.9\textwidth]{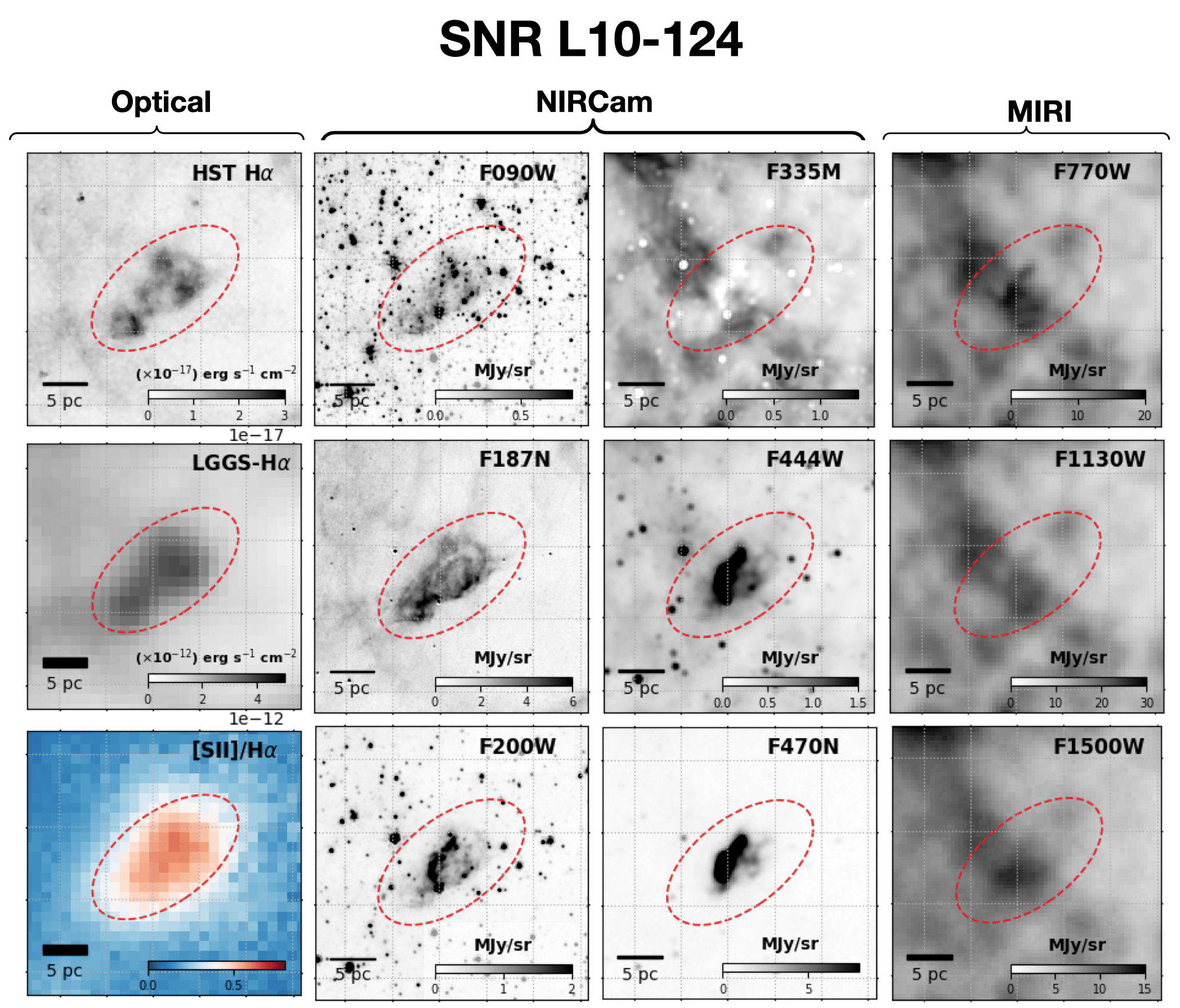}
    \caption{JWST images of the SNR L10-124 in the NGC 604 field. The individual panels show a 7\arcsec\ zoom-in region containing the SNR in optical wavebands from HST and LGGS (first column), and in all 9 JWST filters (remaining columns). The F187N and F335M images have been continuum-subtracted with a scaled continuum from adjacent filters, as described in the text. The dashed ellipse is meant to guide the eye to the SNR's extent, drawn  slightly larger than the extent of the SNR in Pa$\alpha$ image (F187N). }
    \label{fig:l10-124}
\end{figure*}

 \subsection{SNRs in NGC 604} \label{sec:ngc604}

Figure \ref{fig:rgbngc604} shows a three-color image of NGC 604, the largest (visible size of $\sim$450 pc) and the most actively star-forming HII region in M33 \citep[with an average stellar population age of $\sim$4 Myr,][]{NGC604SFR}. The color-image consists of the NIRCam filters F90W, F200W and F470N. As shown in the figure and discussed in Section \ref{sec:obs:jwst}, the NGC 604 mosaic consists of a primary field that covers the full HII region with multiple MIRI and NIRCam filters, and an off-center field that is covered only with the NIRCam filters (see Table \ref{tab:bands} for a list of these filters). Within this mosaic are three known SNRs -- L10-118, L10-124 and LL14-168. The SNR closest to the HII region (at a projected distance of 27.8\arcsec\ from the center) and also the smallest is L10-124. Farther out to the north are L10-118 and LL14-168, at projected distances of about 2.21\arcmin\ and 2.65\arcmin\ respectively. These are the only SNRs in our survey with more than two JWST filter images, and are thus exciting reference sources for the understanding how the spatial morphologies of the SNRs vary from near- to mid-IR.
 \subsubsection{SNR L10-124}
Figure \ref{fig:l10-124} shows a panchromatic zoom-in of the SNR L10-124 in optical, NIRCam and MIRI filters. The SNR is the closest of the three to the center of NGC 604, and also the smallest, with a projected diameter of roughly 4\arcsec$\times$2\arcsec\ (or $\sim$16$\times$8 pc) in the HST images. The SNR is quite bright, visible in almost all the filter images shown in Figure \ref{fig:l10-124}, likely owing to its gas-rich environment visible in Figure \ref{fig:rgbngc604}. 

The first column in Figure \ref{fig:l10-124} indicates the optical morphology of the SNR in a high-resolution HST H$\alpha$ image from archival F658N WFPC2 observations\footnote{\url{https://alasky.cds.unistra.fr/HLA-hips/filter_NII_hips/}}, the lower-resolution LGGS H$\alpha$ images, and the \sii/\ha\ ratio image.  The latter confirms that the full H$\alpha$-visible region is consistent with being a shock-heated plasma. Similar to L10-036, the SNR has a distinct non-spherical, elongated morphology (reminiscent of a pizza-slice). 

The second column shows the morphology of the SNR in the 0.8-2.2 \mum range, with the continuum-subtracted (using F200W), narrowband F187N filter image (containing the Pa-$\alpha$ 1.87 \mum line), the F090W filter (containing the [\ion{S}{3}]$\lambda$9069,9532 doublet, and the $n>7$ Paschen-H series),  and the F200W image (containing the prominent Pa-$\alpha$ and the 2.12 \mum H$_2$ $v$=1$-$0 S(1) lines, among other ro-vibrational lines of H$_2$). The consistent shape across the filters suggests that the observed emission is likely dominated by the hydrogen line series. The consistency is particularly notable between the F658N H$\alpha$ and F187N Pa-$\alpha$ images, with the latter showing a smoother morphology due to lower interstellar extinction in this filter compared to the former. Both F090W and F200W images show traces of the Paschen-H lines, with F200W additionally showing hints of a bright northern arc, which as we show next, is most likely a molecular shock traced by the 2.12 \mum H$_2$ line.

The third column shows the SNR morphology in the 3.1-5.1 \mum range. Similar to the previous column, the filters pick up some stellar emission from old, giant-branch, dusty stars, and some PAH emission in the F335M image (note that the F335M images have Figures \ref{fig:l10-124}-\ref{fig:ll14-168} a subtracted scaled-continuum value, obtained from the flanking F200W and F444W filters). The SNR is visible in emission in all images, and has a distinctly different morphology compared to the optical and Pa-dominated IR images. The IR-emitting region of the SNR is dominated by a bright arc towards the north in F335M/F444W/F470N, similar to F200W. The arc is particularly bright in the F470N filter, which is centered on the rotational H$_2$ $v$=0-0 S(9) transition line\footnote{Some of the P/R branch CO ($v$=1$-$0) ro/vibrational lines also fall within the F470N filter, but these are typically fainter than the H$_2$ line in interstellar shocks \citep[e.g][]{Reach2000, Ray2023}. In ejecta-dominated SNRs like Cas A, lines from freshly formed CO in the ejecta would dominate this filter \citep{Rho2012,Rho2024}, but most of our M33 SNRs are much larger/older.}. The emission from the SNR in this filter is dominated by line emission from H$_2$, as opposed to continuum in the 4-5 \mum window, as evidenced by the significant excess in the F470N:F444W ratio image (Figure \ref{fig:f470n}), with the SNR region showing ratios of 2-6, while the rest of NGC 604 generally shows ratios of $\lesssim$1. Based on this, and the fact that this arc is similarly visible in F200W, which contains the H$_2$ 1-0 S(1) 2.12\mum line, we conclude that the arc structure is likely a molecular shock driven into the surrounding dense environment. This makes sense considering the location of the SNR in NGC 604, one of the most vigorous star-forming regions in the Local Group with significant molecular gas. 

The fourth column shows the MIRI filters covering the wavelength range of 6.5-16.6 \mum. The images reveal the presence of significant amounts of gas in the vicinity of the SNR, traced by PAH-emission and warm dust heated by the ambient radiation field, down to $\sim$2 pc scales. The SNR intersects with a bright filament of gas running northeast to southwest, and a small silhouette of the SNR can be seen along this filament, but the overall SNR morphology seen in the optical or near-IR in the previous columns is not as apparent in these MIRI filters. The bright arc from the NIRCam images however is still visible in F770W, but gradually becomes more muted in F1130W and indistinguishable in F1500W from the surrounding emission. The rotational lines 0-0 S(5) (6.91 \mum) and 0-0 S(4) (8.03 \mum) of H$_2$ occur within the F770W filter bandwidth, so this is likely causing the observed arc \citep{Neufeld2007}, although only future spectroscopy can reveal if there is also some contribution from the 7.7 \mum PAH emission complex. 

Implications of these near-IR and mid-IR observations of the NGC 604 SNR are further discussed in Section \ref{sec:discussion}.

 \subsubsection{SNR L10-118}
Figure \ref{fig:l10-118} shows a similar set of cutout images of L10-118 in all the NIRCam filters, with optical and radio image cutouts as reference to its full structure. This is a larger SNR than L10-124, with a diameter of about 57 pc \citep{White2019}, and as seen in the optical image, much fainter. The SNR has significant north-south asymmetry, with the southern hemisphere significantly brighter in \ha than the northern. The full extent of the SNR is visible in the radio image. 

No detectable emission is present in any of the NIRCam filters, with the exception of F470N. As mentioned before, this filter traces the H$_2$ S(9) line, and the intensity of the filaments in this filter is higher than the continuum-tracing F444W (Figure \ref{fig:l10-118}), implying again the presence of shocked H$_2$. Strikingly, the H$_2$ emission is almost completely anti-correlated with the \ha emission, which is brighter in the south. This is likely due to the ISM being lower density in the south than the north, leading to a faster radiative shock in the south with post-shock temperatures suited for \ha emission, and slower shock in the north where the cooling would be more dominated by infra-red lines from molecules. This is corroborated by the distribution of cold gas shown by the CO(1-0) image and F335M image. Both show an over-dense ridge of ISM in the north coinciding with the shock hemisphere that is bright in F470N.

  \subsubsection{SNR LL14-168}
The third SNR in the NGC 604 field is LL14-168, shown in Figure \ref{fig:ll14-168}. This is the largest of the SNRs in the M33 field, and also one of the largest in M33 with a diameter of about 82 pc \citep{Lee2014, White2019}. Only the southwestern part of the shock is prominent in the \ha image, with no detectable emission in radio above a 3$\sigma$ limit of 12 $\mu$Jy. The narrowband \siiha in Figure \ref{fig:ll14-168} shows ratios somewhat lower than our 0.4 cutoff, but this is most likely due to filter contamination with \ion{N}{2} lines (Section \ref{sec:obs:method}).  Optical spectroscopy by \cite{Long2018} and \cite{SITELLEM33} confirms that \siiha$\gtrsim$0.4 through this arc, consistent with an evolved radiative shock.

The SNR does not show any detectable emission in any of the NIRCam filters, with the exception of F187N which traces Pa$\alpha$ 1.87 \mum. As expected, the F187N-emitting region is roughly the same as the \ha-emitting region, and the ratio of median brightnesses of \ha to F187N is $\sim$8.9, close to the theoretically expected value of 9.1 for a plasma with $T_{\rm e}=10^4$ K, $n_{\rm e}=10^2$ \cmc, assuming Case B recombination \citep{Ferland2017, Liu2024}. Comparison with images from ALMA CO(1-0) and F335M shows that this southwestern shock is interacting with dense ISM. An F335M-emitting arc can be seen in projection in front of the \ha arc of the shock. The CO-emitting gas is also brighter along this arc than within the SNR region. From spatial morphology alone, the F335M and CO-emitting arcs can be interpreted as ISM dispersed and swept up by the SN shock. The fact that the interaction region is only visible in \ha and not in molecular emission (similar to the southern lip of L10-118 shock) implies that the ISM is not dense enough for excitation of H$_2$ ro-vibrational lines \citep{Lehmann2022}.

\begin{figure}
    \centering
    \includegraphics[width=\columnwidth]{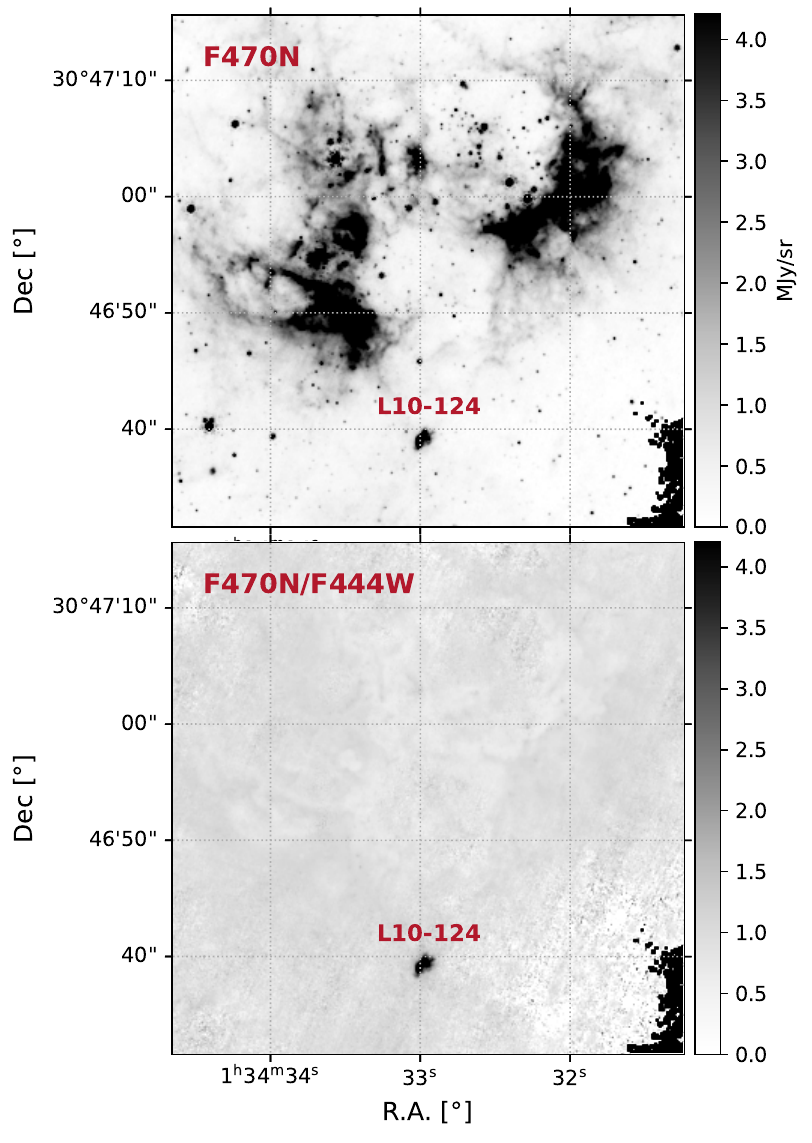}
    \caption{\emph{Top}: The narrowband F470N image of a part of NGC 604, showing the SNR. Image units are in MJy/sr. \emph{Bottom}: Ratio of the narrowband F470N to the broadband F444W filter of the same region. The images confirm that the SNR is exciting H$_2$ gas in the vicinity, given the presence of significant excess emission in the F470N filter over the F444W continuum, in contrast with the rest of NGC 604 where the gas is mostly photoionized (Section \ref{sec:ngc604}). }
    \label{fig:f470n}
\end{figure}
\begin{figure*}
    \centering
    \includegraphics[width=\textwidth]{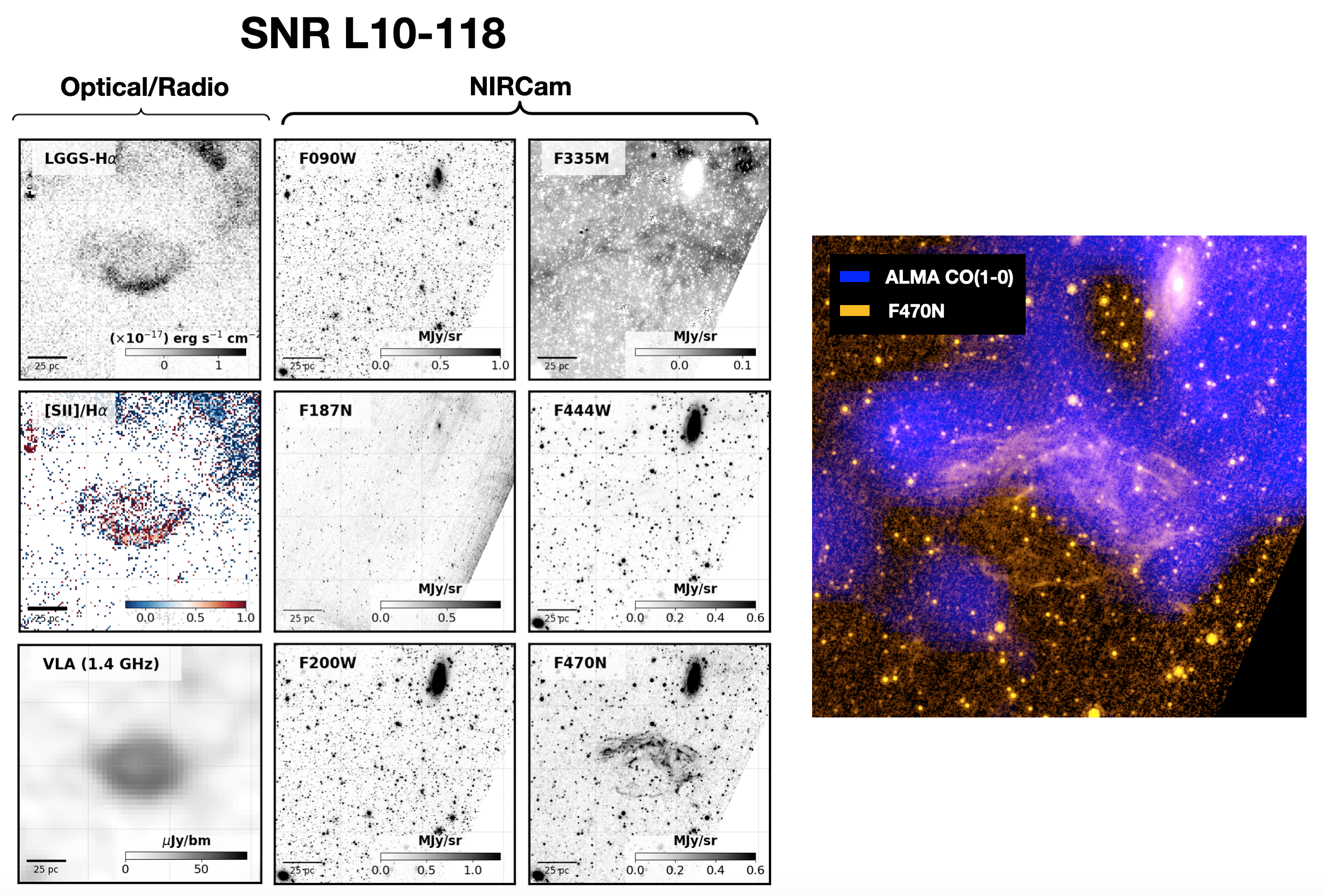}
    \caption{Multipanel images of L10-118 similar to Figure \ref{fig:l10-124} but without MIRI filters. First column consists of the \ha, \siiha and 1.4 GHz radio image that we use as reference for the shock structure. Second and third columns show the NIRCam images, with the filter labeled in the top left corner of each panel, and intensity range in bottom right colorbar. Rightmost cutout is an RGB image (F470N=red+green, ALMA CO(1-0)=blue) of L10-118 and its environs.}
    \label{fig:l10-118}
\end{figure*}
\begin{figure*}
    \centering
    \includegraphics[width=\textwidth]{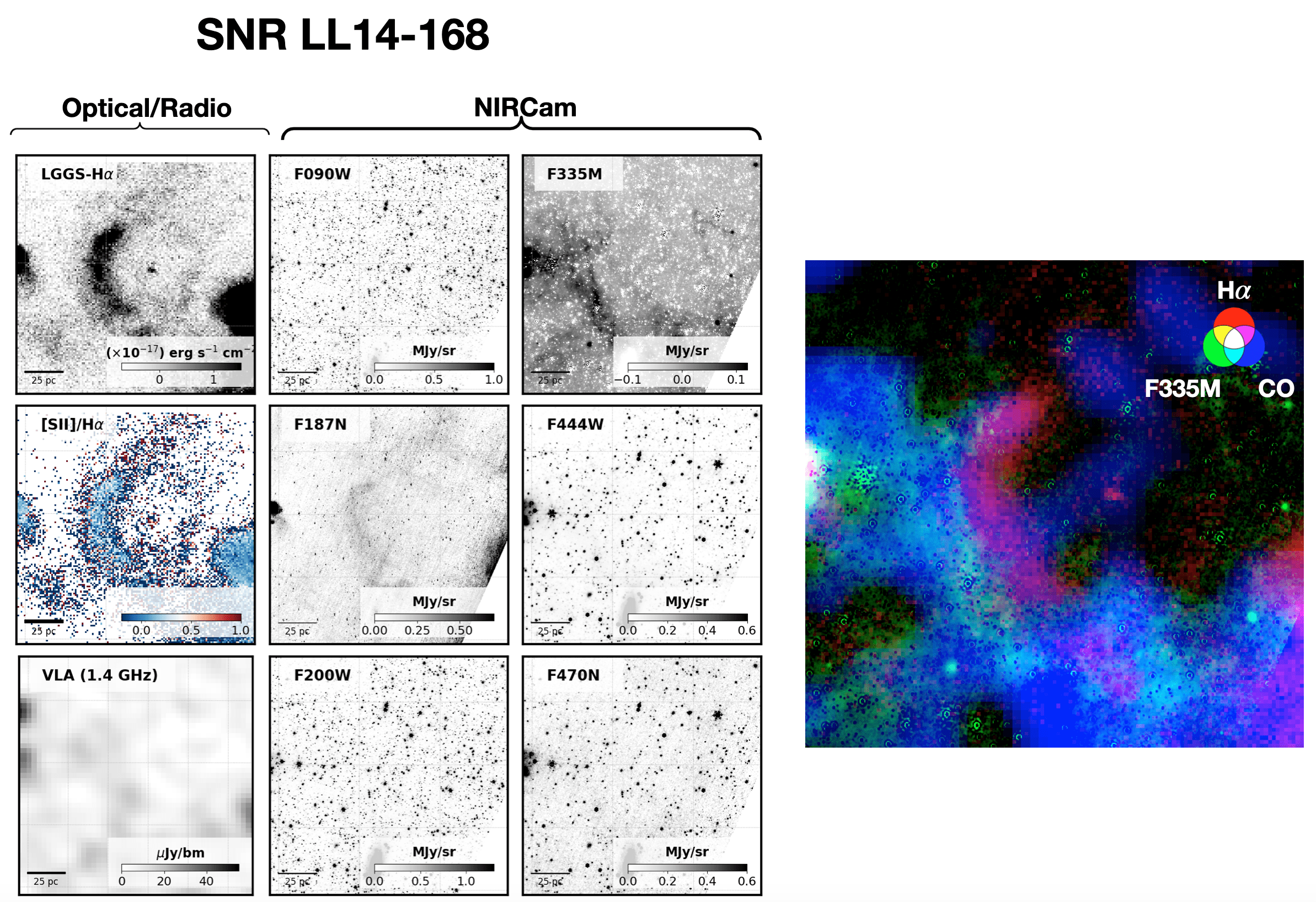}
    \caption{Multipanel images of SNR LL14-168 in the same format as Figure \ref{fig:l10-118}. Left column shows the \ha, \siiha, and radio images, while second and third columns show the NIRCam images. Rightmost is an RGB image (\ha=red, F335=green, ALMA CO(1-0)=blue) of the SNR and its environs.}
    \label{fig:ll14-168}
\end{figure*}

\subsection{SNRs in the North Radial Strip}
\begin{figure*}
    \includegraphics[width=\textwidth]{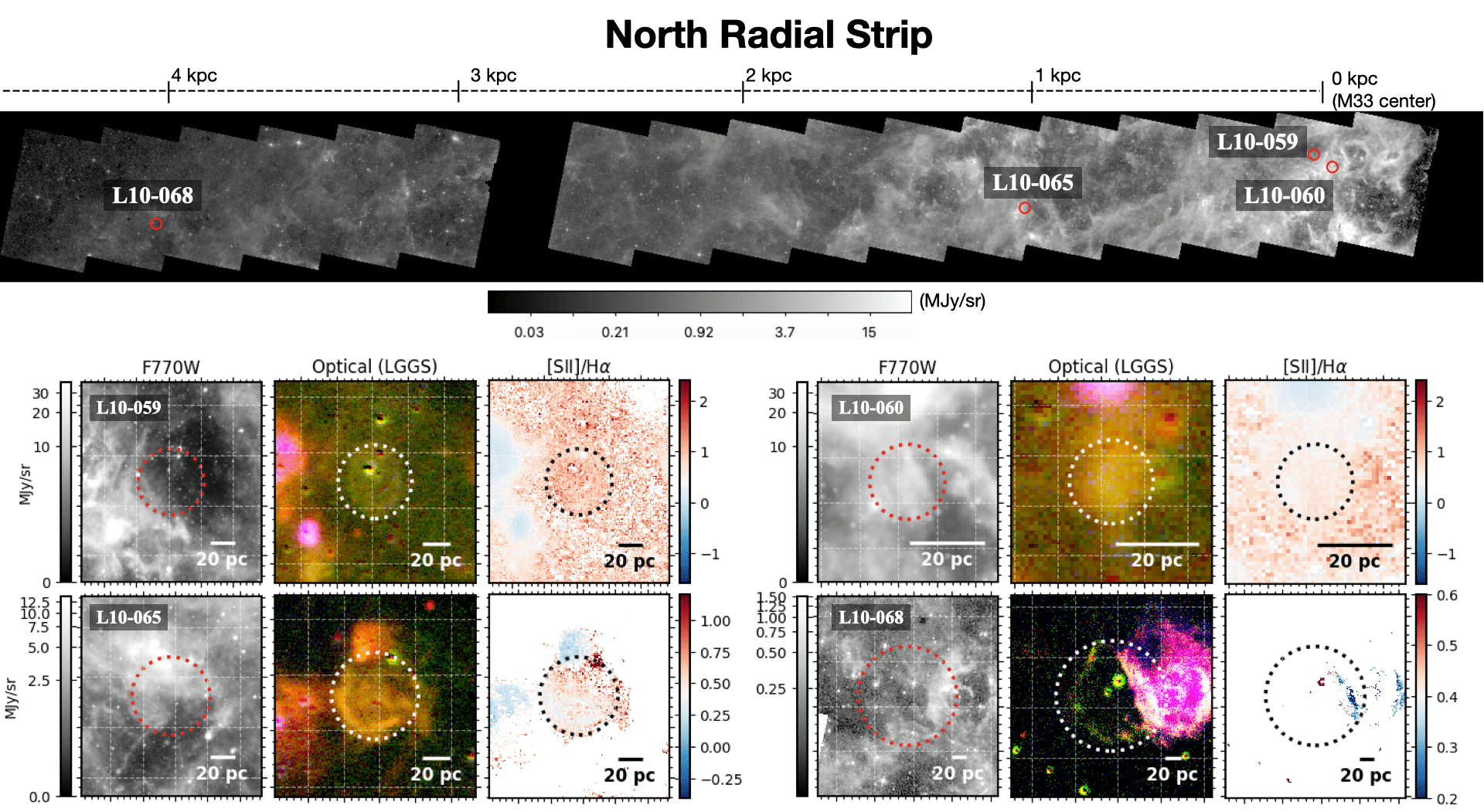}
    \caption{The North Radial Strip and its SNRs. Top image shows the radial strip in grayscale (units in MJy/sr) going from the center of M33 on the right to the outskirts on the left. The four SNRs in the strip are labeled and circled in red. Zoom-ins of the four SNRs are shown in the snapshots below. For each SNR, the left panel is the F770W image in greyscale, the middle panel is the three-color LGGS optical narrowband image same as Figures \ref{fig:brightmirisnrs} and \ref{fig:nircamsnr}, and the right panel is the [\ion{S}{2}]/H$\alpha$ image. Dotted circles are centered on each SNR, and their sizes are 20\% larger than the reported size of each SNR.}
    \label{fig:f770wstrip}
\end{figure*}

The strip is shown in Figure \ref{fig:f770wstrip}, with bright emission likely dominated by the 7.7 \mum PAH feature based on previous Spitzer work \citep[e.g.][]{Smith2007, Calapa2014}. The pc-resolution North Radial Strip ISM shows a complex network of bubbles and filaments, similar to F770W observations in other similar star-forming galaxies \citep[e.g.][]{Leroy2023, Watkins2023}. The emission is brightest on average towards the center of M33, and declines with distance away from it. A more detailed analysis of the PAH emission in this strip will be presented in future work. We find four SNRs located fully within the mosaic. Two SNRs L10-070 and L10-072 had partial overlap (less than half of their optical-emitting regions within the strip), so we omitted them.

Below the image of the North Radial Strip in Figure \ref{fig:f770wstrip}, we also show snapshots of the SNRs in the three-panel plots. The snapshots follow the same pattern as the MIRI and NIRCam snapshots to determine the emitting region in the JWST images: an F770W flux image (in MJy/sr) shown with an arcsinh stretch to balance bright and faint structures, the continuum-subtracted LGGS \ha, \sii and \oiii images to show the optical-emitting nebulae in the region, and \siiha with the color-scale divided at 0.4 to distinguish between photoionized and shock-heated nebulae. Bluer regions denote photo-ionized gas (e.g.\ HII regions) and red denotes shock-heated gas or diffuse ionized gas.   

A possible detection candidate in the North Radial Strip is SNR L10-060, which shows a distinct patch of F770W emission within its optical-emitting region in Figure \ref{fig:f770wstrip}. This is the smallest of the five SNRs in the strip, with a diameter of 17 pc. It also appears in the Center field, and is the only SNR other than SNR L10-036 with a hint of emission in the F335M filter (although the emission is challenging to make out in the glare of the un-subtracted point sources). The shape of the SNR is not entirely clear at the resolution of the LGGS image (the SNR is only $\sim$4\arcsec\ across) and with all the confusing emission around, so it is not clear if the spatial extent of the optical and F770W-emitting regions are exactly the same (like e.g.\ L10-036 or L10-071). A high-resolution continuum-subtracted HST H$\alpha$ or JWST Pa$\alpha$ could confirm whether the IR and optical-emitting regions are the same, and whether the shock is producing the F770W emission.

The other SNRs in the strip are not as clearly identifiable in F770W emission, especially against the bright PAH emission from the surrounding ISM. This is not unexpected, as the PAH emission depends on the radiation field and ISM density for a given dust-to-gas ratio, and all the SNRs are located adjacent to bright star-forming regions (blue patches in the \siiha image) where both quantities tend to be high \citep[e.g.][]{Leroy2023, Sandstrom2023}. The SNRs are also generally larger, with diameters spanning the range of 45 pc for L10-059 to 112 pc for L10-068\footnote{This SNR is quite faint, with a shell barely visible in the saturated cutout in Figure \ref{fig:f770wstrip}, so most of its pixels did not pass our 3$\sigma$-cut used to construct the \siiha images (Section \ref{sec:obs}). The object however was spectroscopically confirmed as an SNR by \cite{Long2018}.}. These SNRs show varying degrees of emission in the vicinity, but none that can be uniquely associated to the optical-emitting region. Some interesting morphologies can still be discerned, such as in the case of L10-065, where the top of the SNR is PAH-rich, the southwest region has a thick arc of emission similar to its optical shell, and the southeast region has little emission, in stark contrast with the rest of the SNR region. Similarly L10-059, which is close to the center and L10-060 (the PAH-detected SNR), also appears in a PAH-void region. Further discussion about the appearance of these SNRs and their interaction with the ambient PAH-rich ISM is in Section \ref{sec:discussion}.

\subsection{Multi-wavelength Properties} \label{sec:res:multiwavelength}
\begin{figure*}
    \centering
    \includegraphics[width=0.9\textwidth]{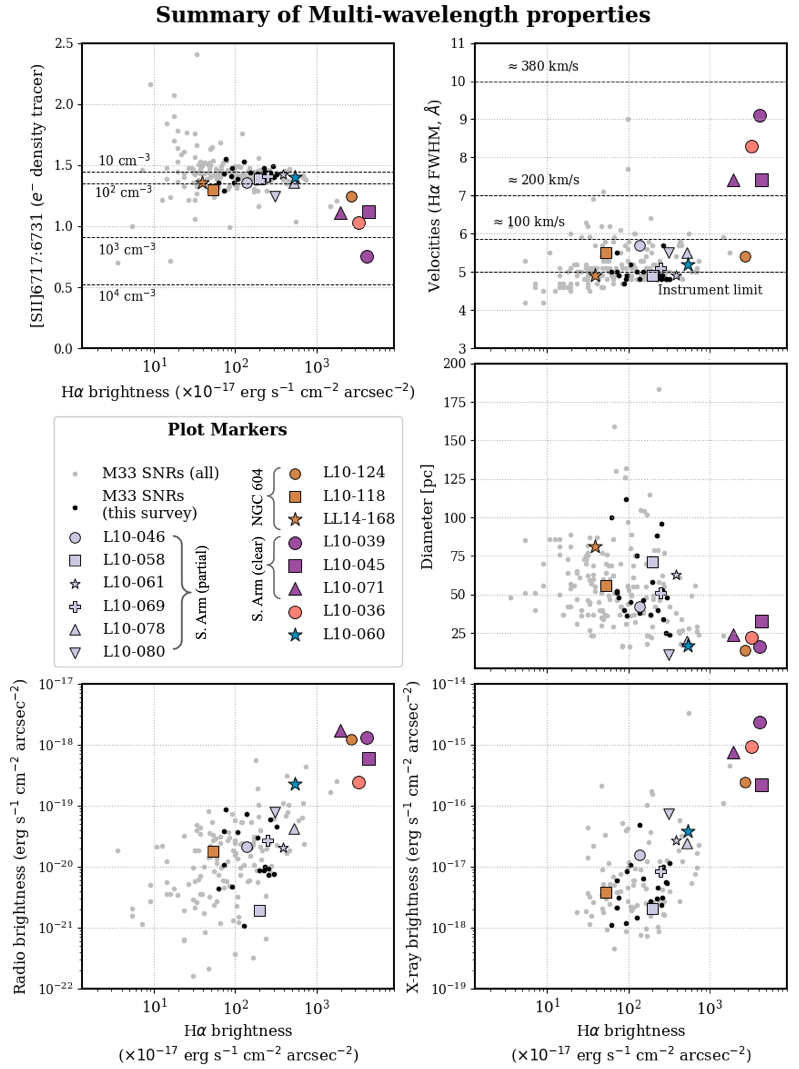}
    \caption{Summary of the optical spectroscopic properties of the SNRs from Long et al (2018). From top left to bottom right, we display the \siidoublet which is related to the ambient density, \ha full-width at half-maximum which relates to the typical shock velocities, diameter, 1.4 GHz radio surface brightness, and the 0.3-2 keV X-ray surface brightness of SNRs. Grey points are the full M33 SNR sample from Long et al (2018), black points are the subset covered by our JWST survey, and colored points denote specific SNRs of interested labeled in the Plot Markers box (Section \ref{sec:res:multiwavelength}). Note that ``S. Arm'' = Southern Arm field.} 
    \label{fig:opticalspectra}
\end{figure*}

We show the multi-wavelength properties of the SNRs in the different JWST fields in Figure \ref{fig:opticalspectra}, as derived from the optical, radio and X-ray datasets introduced in Section \ref{sec:obs}. While a multitude of information is available, we only a display a subset that is most frequently used to assess  shock conditions in extragalactic SNRs --- the density-sensitive \siidoublet line ratio, the characteristic shock velocities from the full-width half-maximum (FWHM) of the H$\alpha$ line, and the diameter estimated by \cite{Long2010} and \cite{Lee2014} using the narrowband optical images from the LGGS survey. We will occasionally refer to the \ha FWHM in this section by its equivalent velocity FWHM in km s$^{-1}$ \citep[similar to]{Long2018}, by subtracting out in quadrature the instrumental limit of 5 \AA\, and an assumed 20 km/s value due to thermal line broadening. We also use published 1.4 GHz VLA radio brightnesses from \cite{White2019} and Chandra-based X-ray brightnesses from \cite{Long2010}, which are both sensitive to the average ambient densities and ages of the SNRs. 


It is clear from Figure \ref{fig:opticalspectra} that one of the distinguishing features of the cleanly detected SNRs compared to the rest of the sample is their higher than average ambient densities. The most direct evidence of this comes from the \siidoublet ratio which is a well-known density diagnostic. The clearly detected SNRs have doublet ratios of 0.75-1.11, which correspond to electron-densities\footnote{We use the \texttt{pyneb} software \citep{pyneb} to convert the \sii doublet ratio values from optical spectroscopy \citep{Long2018} to electron densities for a given plasma temperature} of $\sim$450-2000 \cmc, assuming a plasma temperature of 10$^4$ K \citep[valid for the post-shock region where \sii ionization state is abundant,][]{Allen2008}. The actual temperature of the \sii-emitting zone may still vary depending on other shock parameters. Assuming a factor of 2 variation in temperature, e.g. for a 5$\times$10$^3$K plasma, the \sii doublet values correspond to a lower range of densities ($\approx$330-1440 \cmc), whereas a 2$\times$10$^4$K plasma temperature corresponds to a higher range of densities ($\approx$540-2540 \cmc). Regardless of the temperatures, these densities are much higher than that of the partial and ambiguous MIRI SNRs, which have values more consistent with $<$100 \cmc given their \sii doublet ratios in the range of $\sim$1.3. The scatter in the \sii-doublet values increase as one moves to lower surface brightness SNRs. Note that the \sii-doublet-based densities refer to densities in the extended S$^{+}$ recombination zone of secondary radiative shocks driven into dense cloudlets \citep{Allen2008}. While high \sii-doublet-based densities also imply high preshock densities, estimating the latter will require more detailed comparison between the spectra and radiative shock models \citep[e.g. as done in][]{Dopita2018}, which is outside our scope here. 

Aside from the low \sii ratios, the cleanly-detected SNRs are also distinguished by their smaller sizes, large values of \ha FWHM, and high \ha, radio and X-ray surface brightnesses, compared to the rest of the sample. The smaller sizes of the cleanly-detected SNRs further corroborate the presence of high ambient densities, as do their locations close to dense IR-bright clouds in Figure \ref{fig:miricutout}. If we assume the SNRs are mostly radiative (given their strong \oiii emission), that means the SNRs have at least crossed their cooling radii, which for a turbulent ISM is given by $R_c \approx (\mathrm{32\ pc})n_0^{-0.46}$, where $n_0$ is the average density of the turbulent ambient medium in units of cm$^{-3}$ \citep[e.g.][]{Martizzi2016}. The radii of $\approx$8-16 pc measured for these SNRs therefore implies $n_0 > 5-20$ \cmc. The high values of \ha FWHM indicate the presence of higher-than-average shock velocities\footnote{As discussed in \cite{Points2019}, the few $\times$100 km/s radiative shocks observed in SNRs are likely secondary shocks driven into dense clouds, as these are more likely to reach post-shock temperatures $\lesssim10^5$ K, where the onset of significant cooling occurs. The primary blastwave is likely propagating faster by a factor $\sqrt{\rho_c/\rho_0}$, where $\rho_c$ is the cloud density, $\rho_0$ is the intercloud density, and for clouds, $\rho_c \geq \rho_0$.}, which is a common feature in optical observations of young (few $\times$10$^3$ yr) SNRs \citep[e.g.][]{Fesen2020, Li2021}. Models predict that emission from shocks at these wavelengths increase with both density and velocity \citep[e.g.][]{WhiteLong1991, Allen2008, Sarbadhicary2017}, so the high luminosities at these wavelengths are broadly consistent with the SNRs with high ambient densities, as well as high shock velocities. 

Figure \ref{fig:opticalspectra} also explains why L10-036 is the only visible SNR in the Center field. Similar to the cleanly detected MIRI SNRs (L10-039, L10-045 and L10-071), the multiwavelength properties of L10-036 are quite distinct from the other SNRs as seen in Figure \ref{fig:opticalspectra}, exhibiting much faster radiative shocks (based on the \ha FWHM of 8.2 \AA\ which correspond to shock speeds of about 300 km/s), propagating into high-density ambient clouds (according to the \siidoublet of 1.02, corresponding to electron density of $\approx$658 \cmc) than the other SNRs. Consequently, L10-036 is also among the brightest in \ha, X-ray and radio, and smallest in diameter in our MIRI+NIRCam sample. Note that from the higher-resolution JWST and HST images, we measure the major and minor axes of the SNR as $D_{\rm maj}\approx 3.37$\arcsec\ ($\sim$13.7 pc), and $D_{\rm min}\approx$1.92\arcsec\ ($\sim$7.8 pc), smaller than the previously reported diameters of 18-22 pc from lower-resolution images \citep{Long2010}.

The prominent L10-124 SNR in the NGC 604 field falls in a similar region of the multi-wavelength parameter space in Figure \ref{fig:opticalspectra} as the clearly-detected MIRI SNRs and L10-036. A notable however exception is its \ha FWHM, which is only slightly broader (5.4 \AA) than the instrument limit. The pre-shock density is also somewhat lower, which may indicate that this is SNR is young and not completely radiative throughout. Indeed as noted in \cite{Long2010}, the \oiii from the SNR is mainly in a knot in the southwestern part of the SNR, and smaller in extent than its \ha region. More on this SNR is discussed in Section \ref{sec:discussion}.

The remaining SNRs in our survey -- partially-detected and ambiguous MIRI SNRs, the non-detection NIRCam SNRs, the North Radial Strip SNRs, and SNRs L10-118 and LL14-168 in NGC 604 -- have comparatively unremarkable properties compared to the clearly-detected ones. Their \sii-doublet ratios cluster at the low-density limit of $\sim$1.45, velocities rarely exceeding the instrument resolution or with hints of moderate broadening ($\lesssim$100 km/s), and diameters, radio and X-ray brightness spanning a wide range of values. Two of the SNRs have radio brightness close to the clearly-detected SNRs, but this could be due to contribution from the bright ambient background (L10-070 is near a bright HII region, while L10-060 is the near the bright central region of M33). These properties are also generally shared by the bulk M33 SNR population (not covered in our JWST survey), as seen from the distribution of gray data points in Figure \ref{fig:opticalspectra}.


\subsection{Progenitor masses of Southern Arm SNRs} \label{sec:results:hstsfh}
\begin{figure*}
    \includegraphics[width=\textwidth]{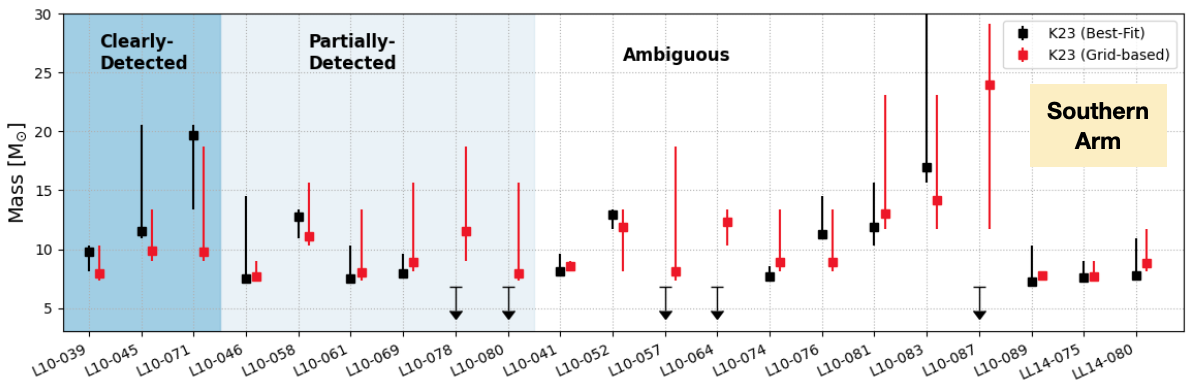}

    \caption{Median and 68th percentile progenitor mass range of our SNRs in the Southern Arm from Koplitz et al (2023, or K23), based on the recent ($<$56 Myr) star-formation histories from PHATTER survey. The two types of masses reported by K23 are shown: 1) ``Best-fit'' mass (black data), which subtracts the star-formation history in an annulus around the SNR, and 2) ``Grid-based'' (red data), referring to masses from the star-formation histories in the 100$\times$100 pc PHATTER SFH cell from Lazzarini et. al (2022). Upper limits indicate SNRs where no significant star-formation within 56 Myr is present in the 50 pc region. Shaded SNRs have detection in one or more JWST filters, as discussed in the text (See Section \ref{sec:results:hstsfh} for details). }
    \label{fig:progmass}
\end{figure*}

Given the distinct range of densities and velocities spanned by the clearly-detected SNRs compared to the other SNRs in the Southern Arm (Section \ref{sec:res:multiwavelength} and Figure \ref{fig:opticalspectra}), it would also be interesting to check if they share a similar distinction in progenitor properties. We check this in Figure \ref{fig:progmass} using the median and 68th percentile progenitor masses of the SNRs in the Southern Arm based on the recent star-formation histories ($<$56 Myr, or $>$7.1 \Msun) around the SNRs, as a proxy for the amount of young star formation in the regions. As a reminder from Section \ref{sec:obs:hstsfh}, these masses were measured by \cite{Koplitz2023} from the age-range of the local stellar population from the PHATTER survey, using stellar age-to-mass relations from single-stellar evolution models. We show both grid-based masses, which are measured from the SFH map of \cite{Lazz2022} at the location of the SNR, and best-fit masses, which were measured by \cite{Koplitz2023} in the $<$50 pc SNR vicinity, after removing the background SFH (see Section \ref{sec:obs:hstsfh} for more details). We denote SNRs of particular interest, i.e. the clearly-detected and partially-detected populations, in two shades of blue.  

Figure \ref{fig:progmass} shows that there is no conspicuous difference in progenitor mass range of the three categories of SNRs in the Southern Arm. For example, SNRs with high-mass progenitors ($>$15 \Msun) can be bright and clearly detected with JWST like the cases of L10-071, as well as ambiguous like the cases of L10-081 and L10-083 (these two are particularly affected by bright background emission). There is also noticeable contrast in progenitor properties within the clearly-detected SNRs. For example L10-039, which is the brightest SNR in the sample at all wavelengths, has a lower mass progenitor of $9.8_{-1.7}^{+0.5}$ \Msun than L10-071 with best-fit mass of $19.7_{-6.3}^{+0.9}$ \Msun. It is worth noting however, that \emph{all} the clearly-detected SNRs -- L10-039, 045, 071 -- have significant recent ($<$56 Myr) star-formation (and thus progenitor masses $>$8 \Msun) in their vicinity. In contrast, of the 18 partially-detected and ambiguous SNRs, 5 have best-fit masses $<$8 \Msun. These SNRs were described by \cite{Koplitz2023} as possible but `weaker' Type Ia candidates, because their grid-based masses in Figure \ref{fig:progmass} still indicates the presence of $<$56 Myr star-formation nearby, just not in their immediate $<$50 pc vicinity. These SNRs, however, could also have been produced by a core-collapse progenitor that originated in a star-formation site farther than 50 pc, and drifted to the current SNR location before exploding. The progenitor could also be from the delayed (50-200 Myr) core-collapse channel predicted by \cite{Zapartas2017} from mergers of low mass ($<$8 \Msun) progenitors.

We can thus conclude that whether the SNRs are clearly or even partially-detected in infrared is mostly independent of the progenitor, and more directly correlated with the ISM conditions and age of the shock (as shown in Figure \ref{fig:opticalspectra}).  This is generally consistent with the observation that most SNRs in galaxies are in their Sedov/post-Sedov stages of evolution \citep{Gordon1999,Badenes2010}, where interaction with the swept-up ISM is primarily driving the shock dynamics and observed emission, instead of the initial ejecta.

\section{Discussion} \label{sec:discussion}
The results above highlight the exciting potential of JWST for studying extragalactic SNRs at near and mid-IR wavelengths. We find a variety of infrared characteristics of SNRs, from small, young objects clearly detected in every filter (e.g L10-124) to more evolved SNRs detected in single emission-line filters (e.g L10-118). The huge observational gains with JWST over \Spit is evident in Figure \ref{fig:jwst-spitzer-comparison}, where clearly it would have been impossible to isolate the mid-IR ($\gtrsim$20 \mum) emission from SNR L10-071 with the \Spit MIPS images alone, or determine whether the emission in the IRAC images for L10-036 is from the SNR itself or unrelated stellar emission. Below we discuss the results from Section \ref{sec:results} in the broader context of infrared observations of SNRs. 

\subsection{Near and mid-infrared detection statistics}
Visual inspection of the SNRs in the sharp JWST images revealed 3 clear and 6 partial detections out of 21 SNRs in the Southern Arm field, giving a clear detection fraction of 14\% and total detection fraction of 43\%. These numbers are roughly consistent with mid-IR detection statistics of SNRs elsewhere in the Local Group. \cite{Matsuura2022} found that out of the 24 known SNRs in the SMC, 5 have clear detections and 6 have possible detections in \Spit IRAC and MIPS images, giving a clear detection fraction of 21\% and total detection fraction of 46\%. Similarly Galactic SNRs have shown detection rates of 20-30\% with IR instruments, although the Galaxy has a greater degree of line-of-sight confusion from overlapping sources \citep[e.g.][]{Saken1992, Reach2006, Chawner2019}. LMC SNRs show a higher detection fraction of about 62\% (29 out of 47 SNRs detectable at 24 \mum) according to the work of \cite{Seok2013}. This could be due to LMC being an external galaxy with lower line of sight confusion than the Galactic midplane. Physical differences in the environments (e.g. densities, metallicities and dust properties) of the LMC SNRs compared to SNRs in other Local Group galaxies could also be responsible for the different detection fractions \citep[e.g. see discussions in][]{Seok2013,Maggi2016, Auchettl2019}.

\subsection{What is driving broadband emission in the clearly-detected SNRs?}

The IR emission in our clearly-detected SNRs is most likely dominated by forward shock interaction with the surrounding ISM. This is supported by the fact that their IR morphologies match the full extent of their optical emission-line structures (Figures \ref{fig:brightmirisnrs}, \ref{fig:nircamsnr}). The SNRs are likely in their Sedov phases given their X-ray spectra and sizes, which indicate swept-up masses $\sim$100-200 \Msun \citep{Long2010, Garofali2017}, and their optical spectra that lack signatures of high-velocity ($>$1000 km/s) ejecta that has been observed in well-known ejecta-dominated SNRs such as Cas A, E0102, and N132D \citep{Long2010, Long2018}.

The presence of dense ambient media around the clearly-detected SNRs is characteristic of other well-known infrared-bright interacting SNRs. For example, N63A, N49 and N103B are among the brightest 24 \mum-emitting SNRs in the LMC \citep{Seok2013}, the smallest ones \citep{Badenes2010}, and have published optical spectroscopy that show \siidoublet of 0.5-1 \citep{Russell1990, Li2021}. The presence of dense ambient gas has also been independently confirmed by interaction signatures with molecular clouds \citep{Oliva1989, Banas1997, Sano2018, Sano2019, Sano2023}. Many of the IR-brightest MW SNRs also share these characteristics \citep[e.g.][]{Reach2006, Pinheiro2011}, and also have independent evidence of interaction with dense nearby clouds  \citep[e.g.][and references therein]{Jiang2010, Slane2015, Kilpatrick2016}. Spectroscopic observations of many of these local SNRs suggest that the F560W emission of the clearly-detected SNRs is likely driven by prominent lines of [\ion{Fe}{2}] 5.34 \mum and H$_2$ S(7) 5.51 \mum, while the emission in F2100W ($\sim$18.7-23 \mum) is produced by continuum emission from collisionally-heated dust  \citep[e.g.][]{Reach2002, Neufeld2007, Hewitt2009}. 

Emission in the wavelength range of our NIRCam filters has also been seen in interacting SNRs, but less commonly than at longer IR wavelengths \citep{Oliva1999, Reach2006, Seok2013, Matsuura2022}. Near-IR spectroscopy of some of these SNRs indicate that the emission in our F335M filter could be produced by the 3.3 \mum PAH C-H stretching band, as seen in the LMC SNR N49 \citep{Seok2012}, and in F444W by prominent spectral lines such as Br$\alpha$ (4.05\mum), Pf$\beta$ (4.65\mum) and several H$_2$ and CO transitions \citep{Oliva1999, Reach2006}. Follow-up IFU spectroscopy of our bright JWST targets will reveal the contribution of the various line and continuum features in detail.

\subsection{L10-080: Candidate for ejecta-dust SNR?} \label{sec:l10-080}
The case of L10-080 (Figure \ref{fig:partialsnrs}, last row) is quite interesting and is a potential candidate for ejecta-dust in our sample. We categorized it as `partially-detected’ because even though there is faint F560W emission that spans the \ha extent of the SNR, the F2100W emission appears to be concentrated mostly at its center. This geometry could be indicative of fresh dust being heated by the reverse shock, while the F560W comes from the forward radiative shock. The SNR is also the smallest in our sample (and likely in M33) with a diameter of about 11 pc. Assuming uniform ISM densities of 1-10 \cmc, ejecta masses between 8-15 \Msun \citep[typical for Type II SNe,][]{Martinez2022}, and steep ejecta profiles with indices $n=10-12$, an SNR will remain ejecta-dominated up to diameters of roughly 7-17 pc \citep{Tang2017}. L10-080 is also one of the SNRs with the highest \siiha values (=1.19) in M33 \citep{Long2018}, which may indicate presence of fast ($>$300 km/s) radiative shocks \citep{Allen2008}, although only moderate broadening (FWHM $\sim$100 km/s) is observed from the \ha line profile (Figure \ref{fig:opticalspectra}). 

Altogether, it is possible that L10-080 is an old ejecta-dominated SNR or an SNR transitioning to the Sedov phase, making the ejecta-dust interpretation feasible. It would be exciting addition to the very limited sample of known SNRs with ejecta dust such as Cas A, 1E02, and pulsar wind nebulae like Crab, Kes75, and G54.1 \citep[see][and references therein]{Williams2017}. Follow-up spectroscopy will be useful to further characterize the source, such as the dust temperature and spectral lines typically associated with ejecta dust (e.g. Ar, Ne). 

\subsection{Evidence of shocks interacting with molecular gas}
Our findings indicate that narrowband MIRI and NIRCam filters can be powerful tools for identifying SNRs interacting with cold, dense clouds in the ISM in external galaxies as far as M31 and M33. Both SNRs L10-118 and L10-124, which are in the NGC-604 field, provide the most direct evidence of this, with significant emission in the F470N filters that contain the H$_2$ and CO lines (Figures \ref{fig:l10-124}$-$\ref{fig:l10-118}). Signatures of dense ISM traced by ALMA CO and F335M at the location of the molecular emission-rich shocks further corroborate the presence of interaction. These are also the farthest SNRs (to our knowledge) where such shocks with molecular emission have been found. Additionally, L10-124 shows evidence of the same molecular shock arc in F200W and F770W, which contain other well known ro/vibrational lines of H$_2$ such as 1-0 S(1) 2.12 \mum, 0-0 S(5) 6.91 \mum, and 0-0 S(4) 8.03 \mum. Bright H$_2$ lines have been observed in several SNRs interacting with molecular clouds \citep[e.g.][]{Reach2000,Andersen2011,Rho2021}, and are typically produced in shocks $<$100 km/s traversing gas clouds with densities of $>$10$^3$ \cmc \citep[e.g.][]{Hollenbach1979, Hollenbach1989, Kaufman1996, Lehmann2020, Lehmann2022, Kristensen2023}. Such molecule-rich shocks may also be present in the other clear and partially-detected SNRs, many of which are in close proximity to dense molecular clouds (Figure \ref{fig:miricutout}). 

These observations could be vital in future for measuring the fraction of SNe whose blastwaves interact with dense gas clouds. This correlation of SNe with dense ISM gas is an important factor driving the outcome of supernova feedback in galaxy evolution \citep[e.g.][]{Iffrig2015,Walch2015,Martizzi2016}. \cite{Sarbadhicary2023} showed that up to 40\% SNRs in M33 are in projection with molecular gas $\gtrsim$1 \Msunpc at $\sim$50 pc resolution, and thus could be interacting. This is consistent with the clear detection fractions quoted previously for the MIRI SNRs, which show that at least 14\% of SNRs in the sample are interacting with the ambient ISM. The fraction may be higher if we include some of the partially-detected SNRs (Figure \ref{fig:partialsnrs}), and some of the ambiguous ones that are not currently producing IR emission, but appear to have left voids in the ISM, possibly due to past interactions (e.g. L10-057, L10-083, Figure \ref{fig:nondetectionmirisnrs}). 

\subsection{Effect of shocks on PAHs}
We briefly comment here about PAH emission from SNRs, which is still a somewhat poorly-understood topic. Only five SNRs in our survey (the North Radial Strip objects in Figure \ref{fig:f770wstrip} and L10-124 in Figure \ref{fig:l10-124}) were observed in the F770W filter. We find a hint of a trend similar to the MIRI SNRs: The smaller SNRs L10-060 and L10-124 appear to show a somewhat brighter patch of emission within their optical-emitting region. L10-124 in particular also shows a clear bright F770W-emitting patch in Figure \ref{fig:l10-124} coincident with its bright molecular shock that we identified in the NIRCam filters (hints of the 3.3 \mum PAH feature are also visible in F335M, Figure \ref{fig:nondetectionnircamsnrs}). While we concluded in Section \ref{sec:ngc604} that this could be from additional H$_2$ rotational line series in the F770W bandwidth, since the shape of the F770W arc is almost the same as in the NIRCam images, some contribution from swept-up PAH-rich gas or pre-shock PAH gas irradiated by the precursor could still exist, as has been seen from spectroscopy and imaging of some young, interacting SNRs \citep{Tappe2006, Neufeld2007, Tappe2012}. 

The larger SNRs such as L10-065 and L10-068 do not show a clear correspondence between their optical structures and the F770W emission, but do seem to have regions of emission ``deficit'' near their centers, compared to the surrounding emission. Morphologically at least, one can argue that these deficit regions are consistent with the shocks destroying PAH-emitting grains. Statistically, larger SNRs tend to be older \citep{Badenes2010, Sarbadhicary2017}, so it is possible the shocks in the larger SNRs have had more time to destroy the pre-existing PAH-emitting grains, while in the smaller objects the PAH is still being processed \citep{Micelotta2010}. Recent JWST studies have confirmed that voids in F770W-emission are ubiquitous in galaxies, and are statistically consistent with structures mainly created by correlated SN and pre-SN feedback \citep{Barnes2023, Watkins2023}. It would not be surprising to see this extend down to the scales of SNRs, although it is difficult to confirm with our sample of size of five. Some random alignment of unrelated F770W-emitting clouds in the ISM along the SNR line-of-sight may also be occurring. Future F770W maps covering a larger number of SNRs can help confirm this dichotomy between older and younger shocks. 

\section{Conclusions}
In this paper, we present the first JWST views of a subset of SNRs in M33 showcasing the detailed near/mid-IR structures of extragalactic SNRs that can be revealed in the sub-arcsecond resolution images. We pulled observations from four regions of M33 that were targeted with NIRCam and MIRI filters during Cycles 1 and 2. A combined 40 SNRs fall in these fields, spanning a variety of sizes and ISM environments. We visually inspect whether these SNRs have complete or even partial detections of their optical morphologies (from LGGS survey) in JWST, and whether there is any correlation between their infrared detections and published multi-wavelength properties. Our analysis reveals the following:
\begin{enumerate}
\item We estimate, by eye, that about 43\% of the SNRs in the Southern Arm field have complete or partial detections of their optical morphologies in the MIRI filters, and only 6.7\% (1 out of 15 detections in the Center field) in NIRCam filters. These detection statistics are similar to near/mid-IR observations of SNRs in the SMC, but midway between those in the LMC and Milky Way. 

\item Three SNRs in the Southern Arm field -- L10-039, L10-045 and L10-071 -- have infrared morphologies nearly identical to their optical morphologies (Figure \ref{fig:brightmirisnrs}). They are also prominent  in broadband HST filters that capture many expected forbidden lines in the optical (e.g. \hb, \oiii). Six SNRs in the Southern Arm field have partial detections (e.g. a single filament in L10-069, part of the shell in L10-058 etc., see Figure \ref{fig:partialsnrs}), while the rest have no detectable emission or emission that is unique from the background (Figure \ref{fig:nondetectionmirisnrs}). 
\item We have one possible candidate for ejecta dust SNR in the MIRI -- L10-080 -- based on its centrally concentrated F2100W (21\mum) emission (Figure \ref{fig:partialsnrs}, last panel). The SNR is also the smallest in our sample (diameter 8 pc), making it possible that it is an old ejecta-dominated SNR transitioning to Sedov phase.

\item The Center field only has 1 clearly-detected SNR -- L10-036 -- clearly visible in F335M and F44W. The prominent ear-shaped morphology extends to the near-IR as well, and is likely shaped by a dense, asymmetrical environment as evidenced from ALMA images. 
\item Of the three SNRs in NGC 604 with images in multiple NIRCam and MIRI filters (Figure \ref{fig:rgbngc604}), L10-118 and L10-124 show significant emission in F470N, signifying excitation of H$_2$ (and possibly some CO) due to interaction with molecular gas visible in ALMA (Figures \ref{fig:l10-124}, \ref{fig:l10-118}). These are the farthest SNRs where such molecule-rich shocks have been discovered. The smallest SNR L10-124 is visible across 0.9-7.7\mum, although with diminishing visibility in MIRI (F1130W, F1500W). The largest SNR LL14-168 ($\sim$80 pc) shows faint emission only in the F187N filter (Pa$\alpha$), and evidence of swept-up ambient ISM in F335M and ALMA CO (Figure \ref{fig:ll14-168}).
\item Of the five SNRs observed with F770W (tracing PAH) in our survey, the smaller SNRs like L10-060 and L10-124 show patches of elevated emission, while the larger SNRs like L10-65 appear to have emission voids within them.  The prominent molecular shock arc of L10-124 is visible in F770W, likely captured by the S(5) and S(4) transition lines, but they are not visible in F1130W and F1500W.

\item Comparison with published multi-wavelength data reveals that our most prominent JWST SNRs (L10-036, L10-039, L10-045, L10-071, and L10-124) are also the ones evolving in densest gas and with the fastest radiative shocks in the survey. This is consistent with their small sizes (about a factor of two smaller on average than the other SNRs), and locations near dense ISM seen in MIRI and (wherever available) ALMA images. This trend is also consistent with observations of interacting SNRs in the Milky Way and Magellanic Clouds.

\item We do not find a strong correlation between the local ($<$50 pc) star-formation history-based progenitor masses of SNRs from \cite{Koplitz2023} and whether they are clearly or partially-detected or undetected in JWST (Section \ref{sec:results:hstsfh}). All three categories of SNRs plausibly span the progenitor mass range of 8 and 30 \Msun, based either on local star-formation histories measured by \cite{Koplitz2023}, or from the M33 star-formation history map of \cite{Lazz2022}. 
\end{enumerate}

Our work helps make the case for mapping the nearest galaxies with JWST to study their SNR population in the coming years. With known distances (leading to accurate luminosities and sizes), decades of multi-wavelength data, and also being just near enough to be well-resolved, these extragalactic infrared SNR surveys can provide value constraints on the statistics of stellar feedback from individual SN explosions. The near/mid-IR filters of JWST cover a treasure trove of spectral lines and continuum windows that can help verify whether SNRs are interacting with molecular clouds, and their role in the dissociation of H$_2$ molecules, PAH molecules, dust, and triggered star-formation. We will pursue follow-up spectroscopy of our M33 SNRs to shed further light on the underlying contributions to their near-to-mid-IR SEDs, and the properties of their preshock medium.

\section{Acknowledgements}
SKS acknowledges useful science discussions with Alberto Bolatto and Ralf Klessen. This research is based on observations made with the NASA/ESA Hubble Space Telescope obtained from the Space Telescope Science Institute, which is operated by the Association of Universities for Research in Astronomy, Inc., under NASA contract NAS 5–26555. These observations are associated with program(s) 12055, 14610. This paper makes use of the following ALMA data: ADS/JAO.ALMA 2018.1.00378.S, 2022.1.00276.S. ALMA is a partnership of ESO (representing its member states), NSF (USA) and NINS (Japan), together with NRC (Canada), NSTC and ASIAA (Taiwan), and KASI (Republic of Korea), in cooperation with the Republic of Chile. The Joint ALMA Observatory is operated by ESO, AUI/NRAO and NAOJ. The National Radio Astronomy Observatory is a facility of the National Science Foundation operated under cooperative agreement by Associated Universities, Inc. This work is based [in part] on observations made with the NASA/ESA/CSA James Webb Space Telescope. The data were obtained from the Mikulski Archive for Space Telescopes at the Space Telescope Science Institute, which is operated by the Association of Universities for Research in Astronomy, Inc., under NASA contract NAS 5-03127 for JWST. These observations are associated with program 2128, 6555, 3436, 2130. Based on observations made with the NASA/ESA Hubble Space Telescope, and obtained from the Hubble Legacy Archive, which is a collaboration between the Space Telescope Science Institute (STScI/NASA), the Space Telescope European Coordinating Facility (ST-ECF/ESAC/ESA) and the Canadian Astronomy Data Centre (CADC/NRC/CSA). This work is based [in part] on observations made with the \Spit Space Telescope, which was operated by the Jet Propulsion Laboratory, California Institute of Technology under a contract with NASA. EWR and JP acknowledge support from the Natural Science and Engineering Research Council Canada, Funding Reference RGPIN-2022-03499 and from the Canadian Space Agency, Funding References 22JWGO1-20, 23JWGO2-A08. The Flatiron Institute is funded by the Simons Foundation. KS acknowledges funding support from grant JWST-GO-02128.004-A. EWK acknowledges support from the Smithsonian Institution as a Submillimeter Array (SMA) Fellow and from grant JWST-GO-03436.007-A. SCOG acknowledges funding from the European Research Council via the ERC Synergy Grant ``ECOGAL'' (project ID 855130) and from the German Excellence Strategy via the Heidelberg Cluster of Excellence ``STRUCTURES'' (EXC 2181 - 390900948).


%

\vspace{5mm}
\facilities{HST, JWST, Spitzer, Mayall, VLA}


\software{\texttt{CASA} \citep{CASA}, \texttt{astropy} \citep{Astropy}, \texttt{matplotlib} \citep{matplotlib}, \texttt{scipy} \citep{scipy}, \texttt{numpy} \citep{numpy}, \texttt{photutils} \citep{photutils}, \texttt{pjpipe} \citep{Williams2024},
\texttt{multicolorfits}
\citep{multicolorfits}.}



\appendix
\section{Individual descriptions of SNRs in the Southern Arm field} \label{app:individualsnrs}
As is conventional in SNR papers, we provide individual descriptions of SNRs. We primarily focus on their JWST images (also HST for the brightest SNRs). Descriptions of the optical emission line, X-ray and radio morphologies have been discussed in detail elsewhere \citep[e.g.][]{Long2010}.

The clearly detected SNRs are shown in Figure \ref{fig:brightmirisnrs}. These include :-
\begin{enumerate}
\item \emph{L10-039} -- Brightest IR SNR in the Southern Arm field, 2nd brightest \ha SNR. In the JWST images (and also in HST), it appears to be a system of an outer and a high-surface brightness inner shock of about 0.63\arcsec (~2.57 pc) diameter. Could be shock interaction with an inner ring of dense material. The morphology is almost spherically symmetric but incomplete, particularly in the northwest region where the rims show discontinuity. Similar loops seen in W49B. 21 \mum however fills the whole region. Larger arcs present as noted by \cite{Long2010} though the possibility of overlapping SNR cannot be ruled out. Located on the edge of a large dust-cleared bubble of $\sim$108$\times$185 pc diameter.

\item \emph{L10-045} -- Also noted for peculiar morphology -- distinct circular SNR, but according to \sii/\ha\ image, appears to be embedded in a larger shock-heated region extending to the south-west in the image. The north-east loop appears to be part of a cavity wall visible in photoionized emission as well as cold molecular gas and dust as noted in \cite{Sarbadhicary2023}. The optical image appears to have a belt of dust obscuring the center of the SNR. This is visible in emission in the JWST image. Bright blue central star in the SNR seen in the optical. Morphology is not as clearly delineated in 21 \mum emission.

\item \emph{L10-071} -- Both 5.6 \mum and 21 \mum trace out the spherical shell of the SNR. Shell is also visible in X-ray as noted by \cite{Long2010}. Brighter on the side of the molecular cloud in all the images, indicating denser material likely originating from the cloud.
\end{enumerate}

The partially detected SNRs are shown in Figure \ref{fig:partialsnrs}. These include,
\begin{enumerate}
\item \emph{L10-046} -- Uniformly filled H$\alpha$-emitting symmetric SNR. We classify this as `partial’ since there is diffuse IR emission present inside the region, including an incomplete arc on the eastern side, but nothing that completely traces out the SNR. 

\item \emph{L10-058} -- Larger SNR with a more elongated shape. The southern region appears to be more shock-dominated with higher \sii/\ha, while the northern half has lower \sii/\ha . Bow-shock feature to the west, with a bright reddish star that was identified as a YSO by \cite{Peltonen2024}, possibly indicating triggered star formation.

\item \emph{L10-061} -- Region of bright \ha\ emission, located about $\sim$37 pc from a $\sim$5 Myr young-star cluster (ID 652) with a bright HII region. There are some bright IR filaments within the SNR aperture but because of the surrounding emission and the unclear morphology of the SNR, it is hard to confirm if the emission is associated with the SNR defined as noted by \cite{Long2010}.

\item \emph{L10-069} -- Partial morphology visible in 5.6+21 \mum, although the rest of the shell is not visible. It is possible that the visible portion is a dense region shocked by gas, although an overdense region of intervening dust cannot be ruled out.

\item \emph{L10-078} -- Partial filament visible in the south, also coinciding with the brightest part of the SNR morphology in the LGGS image. Detection is primarily in 5.6 \mum (hence the green color). Fainter emission is also observed in the northern rim. 

\item \emph{L10-080} --  This is the smallest known SNR in M33. While the entire \ha structure isn't traced out in IR, a faint halo is seen in 5.6 \mum filling the SNR aperture. A bright point source also appears in 5.6 \mum and in 21 \mum, though the latter is offset from the 5.6 \mum by about 4 pc. No distinguishable counterpart or nebula is visible in the HST images. 

\end{enumerate}

\bibliography{manuscript_final}{}
\bibliographystyle{aasjournal}

\end{document}